\documentclass[twocolumn]{aastex631}

\usepackage{amsmath}
\usepackage{amssymb}
\usepackage{multirow}
\usepackage{graphicx} 
\usepackage{hhline}
\usepackage{hyperref}
\hypersetup{
    colorlinks,
    linkcolor={blue},
    citecolor={blue},
    urlcolor={blue}
}

\begin{document}

\title{Cosmological and Astrophysical Parameter Inference from Stacked Galaxy Cluster Profiles Using CAMELS-zoomGZ}

\correspondingauthor{Elena Hernández-Martínez}
\email{elenahernandezmartinez.academia@gmail.com}

\author[0000-0002-1329-9246]{Elena Hernández-Martínez}
\affiliation{Universit\"ats-Sternwarte, Fakult\"at f\"ur Physik, Ludwig-Maximilians-Universit\"at M\"unchen, Scheinerstr. 1, 81679 M\"unchen, Germany}

\author[0000-0002-3185-1540]{Shy Genel}
\affiliation{Center for Computational Astrophysics, Flatiron Institute, 162 5th Avenue, 
New York, NY, 10010, USA}
\affiliation{Columbia Astrophysics Laboratory, Columbia University, 550 West 120th Street, 
New York, NY 10027, USA}

\author[0000-0002-4816-0455]{Francisco Villaescusa-Navarro}
\affiliation{Center for Computational Astrophysics, Flatiron Institute, 162 5th Avenue, 
New York, NY, 10010, USA}
\affiliation{Department of Astrophysical Sciences, Princeton University, 4 Ivy Lane, Princeton, 
NJ 08544 USA}

\author[0000-0001-8867-5026]{Ulrich P. Steinwandel}
\affiliation{Center for Computational Astrophysics, Flatiron Institute, 162 5th Avenue, 
New York, NY, 10010, USA}

\author[0000-0002-2318-3087]{Max E. Lee}
\affiliation{Department of Astronomy, Columbia University, 538 West 120th Street, New York, NY 10027, USA}

\author[0000-0001-8914-8885]{Erwin T. Lau}
\affiliation{Center for Astrophysics | Harvard \& Smithsonian, 60 Garden St, Cambridge, MA 02138, USA}

\author[0000-0002-5151-0006]{David N. Spergel}
\affiliation{Center for Computational Astrophysics, Flatiron Institute, 162 5th Avenue, 
New York, NY, 10010, USA}
\affiliation{Department of Astrophysical Sciences, Princeton University, 4 Ivy Lane, Princeton, 
NJ 08544 USA}

\begin{abstract}

We present a study on the inference of cosmological and astrophysical parameters using stacked galaxy cluster profiles. Utilizing the CAMELS-zoomGZ simulations, we explore how various cluster properties—such as X-ray surface brightness, gas density, temperature, metallicity, and Compton-y profiles—can be used to predict parameters within the 28-dimensional parameter space of the IllustrisTNG model. Through neural networks, we achieve a high correlation coefficient of 0.97 or above for all cosmological parameters, including \(\Omega_{\rm m}\), \(H_0\), and \(\sigma_8\), and over 0.90 for the remaining astrophysical parameters, showcasing the effectiveness of these profiles for parameter inference. We investigate the impact of different radial cuts, with bins ranging from \(0.1R_{200c}\) to \(0.7R_{200c}\), to simulate current observational constraints. Additionally, we perform a noise sensitivity analysis, adding up to 40\% Gaussian noise (corresponding to signal-to-noise ratios as low as 2.5), revealing that key parameters such as \(\Omega_{\rm m}\), \(H_0\), and the IMF slope remain robust even under extreme noise conditions. We also compare the performance of full radial profiles against integrated quantities, finding that profiles generally lead to more accurate parameter inferences. Our results demonstrate that stacked galaxy cluster profiles contain crucial information on both astrophysical processes within groups and clusters and the underlying cosmology of the universe. This underscores their significance for interpreting the complex data expected from next-generation surveys and reveals, for the first time, their potential as a powerful tool for parameter inference.

\end{abstract}


\section{Introduction} 
\label{sec:intro}

Cosmological surveys promise to unlock solutions to longstanding questions about our Universe. Current and upcoming projects, including the Simons Observatory \citep{2019JCAP...02..056A}, CMB-S4 \citep{2016arXiv161002743A}, the Legacy Survey of Space and Time (LSST) at the Vera Rubin Observatory \citep{2019ApJ...873..111I}, Euclid \citep{2022A&A...662A.112E}, the Nancy Grace Roman Space Observatory \citep{2021MNRAS.507.1514E}, e-Rosita \citep{2022A&A...661A...1B}, and the Square Kilometre Array (SKA) \citep{5136190}, aim to explore the cosmos in unprecedented detail across various wavelengths. These surveys aim to probe the fundamental natures of dark matter (DM) and dark energy (DE) and will attempt to resolve existing tensions in cosmological parameters, such as the Hubble parameter ($H_0$) and the structure growth parameter $S_8$.

Central to these efforts is the challenge of interpreting the rich data sets produced by these surveys, which are tainted by uncertainties in baryonic processes, such as feedback of supernovae and active galactic nuclei (AGN). 
These small-scale systematics represent significant sources of uncertainty. Traditional methods of dealing with these uncertainties involve discarding affected small-scale data, as seen in weak lensing analyses from the Dark Energy Survey (DES) \citep{PhysRevD.105.023515}, the Kilo-Degree Survey (KiDs) \citep{2023A&A...679A.133L}, and HyperSuprimeCam (HSC) \citep{2023PhRvD.108l3518L}. However, this results in a loss of valuable information, highlighting the need for more advanced modeling techniques.

Simultaneously, the standard model of cosmology, $\Lambda$CDM, provides a robust framework to describe the Universe’s evolution. It incorporates critical parameters like $\Omega_{\rm m}$, the matter density parameter, whose accurate measurement is essential for advancing our understanding of cosmic structure formation and evolution. Traditional statistical methods for parameter inference, such as the analysis of redshift-space distortions \citep{1987MNRAS.227....1K, 1977ApJ...212L...3S, 2020ApJ...897...17T} and pairwise velocities \citep{1994ApJ...437L..51C, 2015A&A...583A..52M}, have been instrumental but often sub-optimal due to their reliance on summary statistics like the power spectrum, which do not fully capture the non-Gaussian nature of cosmic fields.

Machine learning (ML) emerges as a transformative approach to overcome these limitations. ML techniques excel at handling complex, high-dimensional data sets, offering a way to perform likelihood-free inference and directly learn from raw data without compressing it into summary statistics. Neural networks, for instance, can be trained on data produced by hydrodynamic simulations to predict cosmological parameters (while marginalizing over baryonic effects), achieving high precision even in the presence of baryonic effects. 

The Cosmology and Astrophysics with Machine Learning Simulations (CAMELS) \citep{2021ApJ...915...71V} project exemplifies the potential of ML in cosmology. CAMELS uses thousands of state-of-the-art hydrodynamic simulations to train models that can extract intricate patterns from large-scale structure data. These simulations include various implementations of galaxy formation and evolution physics, utilizing different codes and subgrid physics such as IllustrisTNG \citep{2017MNRAS.465.3291W, 2018MNRAS.473.4077P}, Magneticum \citep{2016MNRAS.463.1797D, 2024A&A...687A.253H}, SIMBA \citep{2019MNRAS.486.2827D} and Astrid \citep{2022MNRAS.513..670N, 2022MNRAS.512.3703B}. The diversity in simulation techniques aims to provide a comprehensive training set for ML models, with the goal of enabling them to generalize well to real observational data. 

The CAMELS project has demonstrated the power of ML to infer cosmological parameters directly from the properties of collapsed structures. For instance, by training neural networks on galaxy properties and catalogs from simulations, researchers have demonstrated promising precision in constraining $\Omega_{\rm m}$ \citep{2022ApJ...929..132V,2023ApJ...952...69D, 2023ApJ...954..125E, 2023ApJ...944...27S, 2023arXiv231015234D}. This approach leverages the fact that galaxy properties are influenced by cosmological parameters in ways that are distinct from baryonic processes, allowing ML models to disentangle these effects. Additionally, ML techniques have been applied to weak lensing maps 
\citep{2017arXiv170705167S, PhysRevD.100.063514, 2019MNRAS.490.1843R, PhysRevD.102.123506}, 2D maps from state-of-the-art hydrodynamical simulations \citep{2021arXiv210909747V, 2021arXiv210910360V}, and other data types, consistently outperforming traditional methods.

Recently, the CAMELS simulation set has been extended with CAMELS-zoomGZ \footnote{https://zoomgz.readthedocs.io/}, a set of new galaxy groups to cluster scale zoom-in simulations that span a wide range of astrophysical and cosmological parameters in the IllustrisTNG model's parameter space \citep{2024ApJ...968...11L}. This suite is designed to train the CARPoolGP emulator \footnote{https://carpoolgp.readthedocs.io/}, to predict integrated as well as radial halo quantities throughout the entire parameter space, extending the CAMELS experiments at the high-mass end of the mass function. The use of this simulation suite, together with the CARPoolGP emulator, allows for detailed analysis of galaxy cluster quantities, enabling the extraction of information on both astrophysical and cosmological parameters from massive virialized structures and providing a deeper understanding of the complex interplay between cosmological and astrophysical processes in galaxy clusters.

On the observational side, the study of galaxy clusters and galaxy cluster profiles has seen significant progress, with current techniques enabling detailed measurements across various wavelengths. X-ray observations, primarily from telescopes like {\it Chandra} and {\it XMM-Newton}, have allowed precise mapping of the intracluster medium (ICM), revealing temperature and density profiles that offer insight into the gravitational potential of clusters \citep{2006ApJ...640..691V, 2009A&A...498..361P, 2013SSRv..177..119E, 2014Natur.515...85Z}. In the radio band, the Sunyaev-Zel’dovich (SZ) effect is being measured with high sensitivity using instruments such as the Atacama Cosmology Telescope (ACT) and the South Pole Telescope (SPT), providing an independent probe of the ICM pressure distribution \citep{2012ApJ...758...75B, 2013JCAP...07..008H, 2015ApJS..216...27B, 2019ApJ...872..170R, 2021ApJS..253....3H}. Optical and infrared surveys, like those from the Dark Energy Survey (DES) and the upcoming Vera C. Rubin Observatory, offer complementary views of galaxy distributions and lensing profiles \citep{2012ApJ...758...75B, 2018MNRAS.481.1149Z, 2018ARA&A..56..393M, 2019ApJ...873..111I, 2021ApJS..254...24S, 2022PhRvD.105b3515S}. Over the next few years, missions like {\it eROSITA} and {\it Euclid} are expected to provide deeper insights, particularly in the mass and redshift ranges where uncertainties are still significant, improving constraints on both astrophysical processes and cosmological parameters.

In this work, we leverage this new suite of simulations to focus on extracting detailed information from simulated stacked galaxy cluster profiles. Our analysis reveals that stacked galaxy cluster profiles allow us to infer all cosmological as well as all astrophysical parameters of the model with the highest accuracy ever achieved, enhancing our ability to interpret observations from upcoming cosmological surveys. 

The paper is organized as follows: Sec. \ref{sec:data} details the simulations and datasets used in this study, focusing on the IllustrisTNG galaxy formation model and the extensive parameter space covered by our new suite of zoom-in simulations. In Sec. \ref{sec:methodology}, we outline the methodology, including the neural network architecture, training procedures, and the performance metrics employed to assess model accuracy. The results of our parameter inference experiments are presented in Sec. \ref{sec:results}, where we explore the influence of different profile types, mass bins, noise levels, and radial cuts on the accuracy of cosmological and astrophysical parameter estimation. We also discuss the effectiveness of using full cluster profiles versus integrated quantities for parameter inference. Finally, we conclude by summarizing our key results and suggesting directions for future research in Sec. \ref{sec:conclusions}.

\section{Data} 
\label{sec:data}

\subsection{Simulations}

The core CAMELS suite includes over 5,516 cosmological hydrodynamical simulations and 5,164 N-body counterparts, each within a comoving volume of (25 Mpc/h)³. They span multiple galaxy formation models and cover a wide range of astrophysical and cosmological parameter spaces \citep{2021ApJ...915...71V}.

For our study, we focus on the IllustrisTNG galaxy formation model, which builds upon the original Illustris model and utilizes the AREPO Tree-PM moving mesh code \citep{2010MNRAS.401..791S}. The IllustrisTNG model includes five cosmological parameters as well as parameters for supernova and AGN feedback, essential for accurately modeling galaxy formation and evolution \citep{2003MNRAS.339..289S, 2013MNRAS.436.3031V,  2017MNRAS.465.3291W, 2018MNRAS.473.4077P}. While the flagship CAMELS suite parameterizes the IllustrisTNG galaxy formation model with only four astrophysical parameters corresponding to supernova and AGN feedback, more recently \citet{2023ApJ...959..136N}  extended this set to include 2048 simulations sampled from a Sobol sequence \citep{Sobol1967OnTD} over a 28 dimensional astrophysical and cosmological parameter space in the IllustrisTNG model (TNG-SB28), presented in Table \ref{tab:illustrisparams}. This work explores the same 28-dimensional parameter space as in \citet{2023ApJ...959..136N}.

Recently, \cite{2024ApJ...968...11L} introduced a new set of zoom-in simulations, CAMELS-zoomGZ, which spans the full 28-dimensional parameter space of TNG-SB28, with a focus on the high-mass end of the halo mass function. This simulation set opens an avenue for parameter inference across the entire parameter space of the IllustrisTNG model, specifically using galaxy groups and clusters. In this paper, we provide a brief overview of the creation process of this simulation set, with detailed explanations available in \cite{2024ApJ...968...11L}.

The set was designed to optimize sampling within the high-dimensional parameter space of the model, reducing cosmic variance in the process. The approach combines the CARPoolGP method with an active learning procedure. This strategy involves creating ``base" and ``surrogate" simulations that are closely related. The ``base" simulations are zoom-in simulations of galaxy clusters and groups, distributed across the IllustrisTNG-model parameter space \(\theta_i\). Within this parameter space, specific ``parameter islands" \(\theta_S\) are defined. The ``surrogate" simulations are initialized with the same phase of initial fluctuations as the ``base" simulations but are located within these parameter islands \(\theta_S\), maintaining correlated sample variance. In other words, the base and surrogate simulations are built by performing a zoom-in of the same halo at two separate points in parameter space - one at a unique parameter space location (base), and one on a parameter island (surrogate). This leads to a reduction of the ``cosmic variance" between base and surrogate. At the same time as each parameter island contains multiple ``surrogate" simulations, this approach provides a reduced-variance estimate at those locations, thus reducing the overall variance of the simulation set. CARPoolGP makes use of the ``base" and ``surrogates" to perform emulations of the average, through a gaussian process with a modified covariance matrix to account for correlations between these ``base" and ``surrogate" simulations.

To determine the optimal locations in parameter space for running simulations and reducing predictive covariance, an active learning procedure is employed. This reduction in predictive covariance is trained and tested on a specific quantity—in this case, the Compton-y parameter (\(Y_{200,c}\)), chosen for its role as a proxy for halo masses and its representation of the thermodynamic properties of halo gas. The application of this method produced a suite of 768 galaxy group and cluster simulations with masses between \(10^{13}\) and \(10^{14.5} M_{\odot}\), allowing the inclusion of high-mass halos in the simulation suite, and thus providing improved mass coverage across the entire range of parameter space.

\begin{table*}
\centering
\begin{tabular}{|p{0.30\linewidth}|p{0.6\linewidth}|}
\hline
 Parameter     &   Meaning \\
 \hline\hline
$\Omega_{\rm m}$ & Ratio of the matter density of the universe to the critical density. \\
$\sigma_8$   &  Amplitude of matter fluctuations on a scale of  8 Mpc. \\
WindEnergyIn1e51erg (ASN1) & Galactic wind energy (in units of $10^{51}$ ergs) injected into galactic winds by supernova feedback. \\
RadioFeedbackFactor (AAGN1) & Efficiency of energy injection from AGN in their kinetic mode into the surrounding gas. \\
VariableWindVelFactor (ASN2) & Controls wind velocity scaling according to local galaxy conditions, particularly setting the speed 
per unit of star formation.\\
RadioFeedbackReiorientationFactor (AAGN2) & Determines the speed and frequency of feedback injection events in radio (kinetic) mode. 
\\
$\Omega_b$ & Ratio of the density of baryonic matter to the critical density of the universe.\\
$H_0$ & Hubble parameter, i.e. current rate of expansion of the universe\\
$n_s$ & Initial power spectrum spectral index.\\
MaxSfrTimescale ($t_{\text{SFR}}$) & The timescale over which gas is converted into stars at the star-formation density threshold.\\
FactorForSofterEQS (EQS) & Adjustment to the effective equation of state for gas--see \cite{2003MNRAS.339..289S}--in star-forming regions.
\\
IMFslope (IMF) & Slope of the Initial Mass Function (IMF), which describes the distribution of masses for a population of stars at the time of their formation. \\
SNII\_MinMass\_Msun (SNII) & Minimum stellar mass, in solar masses, required for a star to end its life as a Type II supernova (SNII). \\
ThermalWindFraction (W1) & Proportion of the total energy from stellar feedback that is used to heat the gas and drive thermal wind (injected as thermal feedback).\\
VariableWindSpecMomentum (W2) & Regulates how the speed 
of galactic winds changes due to stellar feedback.
\\
WindFreeTravelDensFac (W3) & Controls wind particle travel distance by setting a density threshold, determining how far they move before recoupling 
, affecting galactic wind propagation through the ISM and CGM.\\
MinWindVel (W4) & The minimum velocity imparted to gas particles that are ejected as part of galactic winds. \\
WindEnergyReductionFactor (W5) & The fraction by which the galactic wind energy budget, via the mass-loading, is reduced at higher metallicities.\\
WindEnergyReductionMetallicity (W6)  & Adjusts the typical metallicity at which the galactic wind energy budget will be reduced. 
\\
WindEnergyReductionExponent (W7) & Defines the exponent in the power-law relationship between wind energy and gas metallicity, influencing how strongly metallicity affects the wind energy. \\
WindDumpFactor (W8) & Represents the fraction of metals that are not fully ejected out into the galactic wind, but instead get deposited into nearby star-forming cells before the actual wind ejection. 
\\
SeedBlackHoleMass (BH1) & Initial mass of seed black holes introduced in the simulation to represent the formation of supermassive black holes in galaxies.\\
BlackHoleAccretionFactor (BH2) & Rate at which black holes accrete gas, expressed as a Bondi rate multiplier, affecting black hole growth and AGN feedback processes. \\
BlackHoleEddingtonFactor (BH3) & Normalization factor for the limiting Eddington accretion rate for the accretion oto SMBH. In some cases it may allow super-Eddington accretion scenarios in the simulation. \\
BlackHoleFeedbackFactor (BH4) & Scales the energy released by an accreting black hole to influence its environment through thermal energy, kinetic outflows, or radiation.\\
BlackHoleRadiativeEfficiency (BH5) & Refers to the fraction of accreted gas mass rest energy converted into radiation, affecting AGN luminosity and its environmental feedback.\\
QuasarThreshold (Q1) & Eddington ratio threshold for classifying a black hole in quasar mode feedback, distinguishing AGN feedback modes by accretion rate.\\
QuasarThresholdPower (Q2) &  Sets the power-law dependency of the threshold accretion rate, influencing the steepness of the transition between AGN feedback modes. \\
\hline
\end{tabular}
\caption{Parameters of the IllustrisTNG model.}
\label{tab:illustrisparams}
\end{table*}

\subsection{Emulating Galaxy Cluster Profiles}

The key advantage of CARPoolGP in our study is its ability to utilize the aforementioned simulations to generate low-variance emulations of the desired summary statistics. By leveraging the built-in correlations between ``base" and ``surrogate" samples at different locations in the parameter space, CARPoolGP can produce new averaged samples, known as emulations, at any point within the IllstrisTNG-model parameter space.

In this work, we aim to investigate the information contained in averaged or stacked galaxy cluster profiles. Specifically, we focus on five fundamental gaseous properties in galaxy clusters: X-ray surface brightness (SBx) in the 0.5–2 keV photon energy band, gas density (\(\rho_g\)), gas temperature (T), metallicity (Z), and the Compton-y profile.

We began by extracting the profiles for each galaxy cluster in the CARPOOL-zoomGZ set. The temperature, metallicity, and gas density profiles were derived from the SUBFIND outputs of the simulations, using 3D spherical radial bins evenly spaced in log space up to \(R_{200,c}\). The surface brightness and Compton-y parameters were also calculated using SUBFIND outputs, but these were processed as projected quantities due to their inherently two-dimensional nature.

After extracting the profiles for each group and cluster in the simulation suite, we employed the CARPoolGP emulator to create averaged galaxy cluster profiles of each type across various locations in the IllustrusTNG model parameter space. By using the CARPoolGP emulator, we are able to generate a large number of averaged galaxy cluster profiles and cover any point in IllustrisTNG model parameter space within the upper and lower limits set by the original simulations set—something that would be computationally prohibitive through simulations alone.

\section{Methodology} 
\label{sec:methodology}

\subsection{Architecture and Loss Function}

In this work, we train neural networks to perform inference on the values of all cosmological and astrophysical parameters in the IllustrisTNG model from the emulated groups and cluster observables mentioned above. 
Our architecture comprises multiple fully connected layer blocks, each containing a fully connected layer, followed by a LeakyReLU activation layer with a slope of 0.2 and a dropout layer, where the dropout rate is a hyperparameter. The final layer of the architecture is a fully connected layer.

Our neural network is designed to compute key statistical moments, such as the mean and variance, of the marginal posterior distributions for each parameter, without making any assumption about the shape of the posterior. 

The input to our models is a 1D vector containing the values of an averaged galaxy cluster profile for each radial bin. The model output \(2N_{\text{params}}\) values, where \(N_{\text{params}}\) is the number of parameters considered for inference. For each parameter \(i\), the models return its marginal posterior mean (\(\mu_i\)) and standard deviation (\(\sigma_i\)). This is achieved by minimizing the following loss function:

\begin{equation}
\begin{aligned}
\mathcal{L}= & \sum_{i=1}^6 \log \left(\sum_{j \in \text { batch }}\left(\theta_{i, j}-\mu_{i, j}\right)^2\right) \\
+\quad & \sum_{i=1}^6 \log \left(\sum_{j \in \text { batch }}\left(\left(\theta_{i, j}-\mu_{i, j}\right)^2-\sigma_{i, j}^2\right)^2\right)
\end{aligned}
\end{equation}

This loss function ensures that \(\mu_i\) and \(\sigma_i\) represent the parameter's posterior mean and standard deviation, as described in \cite{2020arXiv201105991J} and  \cite{2021arXiv210910360V}.

The training process uses simulated data \(\{x_i, \theta_i\}\), where the parameters \(\theta_i\) are sampled from prior distributions. The networks learn to map the input data \(x\) to the moments of the posterior distribution. During inference, the trained networks can rapidly evaluate the moments for newly observed data, eliminating the need for extensive sampling or density estimation, thereby streamlining the process of parameter inference in complex cosmological models.

\subsection{Input Data}

The one-dimensional vector that we will use as input can be composed of only one or several concatenated average galaxy cluster radial profiles of various types, extending up to \(r_{200c}\) for a specific point in the IllustrisTNG model parameter space. The quantities considered include radial temperature, metallicity, gas density, X-ray surface brightness, and the Compton-y profile. The first three profile types are divided into 30 bins each, while the latter two consist of 29 bins. Therefore, the input size ranges from 29 to 148 bins, depending on the combination of profiles used.

We analyze the concatenated emulated profiles to assess their impact on the inference of the 28 parameters of the IllustrisTNG model presented in Table \ref{tab:illustrisparams}. Additionally, we investigate the inference performance when using each profile type individually.

To investigate the effect of noise on the profiles, we generated a set of noisy profiles by adding random Gaussian noise to each bin, with noise levels of 10\%, 20\%, 30\%, and 40\% of the bin signal.

To determine where the majority of the information resides, we created a set of profiles by progressively excluding the outer regions, ranging from 0.7\(R_{200c}\) to 0.1\(R_{200c}\). Similarly, we cut out the inner regions from 0.1\(R_{200c}\) to 0.27\(R_{200c}\) to analyze the impact of the outer regions on the inference.

Additionally, we also created integrated values from the profiles to compare the inference performance of the full averaged radial profiles with that of the integrated quantities, providing insights into the relative effectiveness of each approach for different parameters.

\subsection{Training procedure and optimization}

We trained our model on emulated averaged galaxy cluster profiles. We examined the effect of the size of the training set on the inference and concluded that employing 300,000 averaged profiles when using one profile type yields a noticeable improvement in accuracy, but it is more convenient and computationally efficient to use 30,000 averaged profiles of each type when using the five profiles types altogether (see Sec. \ref{sec:A} of the appendix). These profiles were consistently divided such that 70\% were used for training, and 15\% were allocated for both validation and testing respectively, with the samples randomly drawn accross the parameter space. 

We trained the models using the specified architecture for 1000 epochs, employing the ADAM optimizer \citep{2014arXiv1412.6980K} to perform gradient descent, with a batch size of 256 samples. Hyperparameter optimization was conducted using the Optuna package \citep{2019arXiv190710902A}, which employs Bayesian optimization with the Tree Parzen Estimator (TPE) \citep{inproceedings}. The optimized hyperparameters included the number of layers, the number of neurons in the fully connected layers, the learning rate, the weight decay, and the dropout rate. This process involved at least 1000 trials, with Optuna directed to minimize the validation loss. The validation loss was computed using an early-stopping scheme to ensure only the model with the minimum validation error was saved. The selected model was then used for subsequent testing.

\subsection{Performance Metrics}
\label{subsec:3.4}
In this study, we employ four distinct metrics to assess the accuracy and precision of our models. To apply these metrics, we consider a 1-dimensional vector \(i\) of the size of the testing set, where \(\theta_i\) denotes the value of the parameter in question, \(\mu_i\) is the posterior mean predicted by our network for that parameter, and \(\sigma_i\) is predicted standard deviation. The four metrics used are:\\

\begin{enumerate}
    \item {\bf Root Mean Squared Error (RMSE):}

  RMSE quantifies the power of our inference by measuring the average magnitude of errors between predicted and true values. It provides a single number that describes how close the predictions are to the actual values.

  \[ RMSE = \sqrt{\frac{1}{N} \sum_{i=1}^{N} (\theta_i - \mu_i)^2} \]

  \begin{itemize}
      \item {\bf Low RMSE:} Indicates that the predictions are close to the actual values, signifying higher accuracy of the model.
      \item {\bf High RMSE:} Indicates larger errors between predictions and actual values, signifying lower accuracy.
  \end{itemize}

  RMSE is particularly useful in contexts where the magnitude of the error is important and provides an absolute measure of fit.

\item {\bf Mean Relative Uncertainty ($\epsilon$):} 

  The mean relative error tells us about the precision of the model by evaluating the relative size of the standard deviation of the prediction compared to the predicted means. 

  \[ \epsilon_i = \frac{1}{N} \sum_{i=1}^{N} \frac{\sigma_i}{\mu_i} \]

  \begin{itemize}
      \item {\bf Low $\epsilon$:} Indicates higher precision, meaning the predicted values have smaller relative uncertainties.
      \item {\bf High $\epsilon$:} Indicates lower precision, meaning the predicted values have larger relative uncertainties.
  \end{itemize}

  This metric helps us understand the precision of the predictions, although it does not provide information about their accuracy.

\item {\bf Correlation Coefficient ($r$):}

  The correlation coefficient measures the strength and direction of the linear relationship between the predicted and actual values.

$$ r(\theta, \mu)  = \frac{\text{cov}(\theta, \mu)}{\sigma_{\theta} \sigma_{\mu}}$$

  \begin{itemize}
      \item {\bf $r$ close to 1:} Indicates a strong positive linear relationship between predicted and actual values, meaning the predicted values closely follow the trend of the real values.
      \item {\bf $r$ close to -1:} Indicates a strong negative linear relationship between predicted and actual values, indicating that the predicted values follow the opposite trend of the real values.
      \item {\bf $r$ close to 0:} Indicates little to no linear relationship between predicted and actual values, suggesting the predicted values are scattered randomly and do not align with the trend of the real values.
  \end{itemize}

  The correlation coefficient is useful for understanding the degree to which the predictions follow the trend of the actual data.

\item {\bf Reduced Chi-Squared ($\chi ^2$):}

  Reduced Chi-Squared provides a measure of how accurate the predicted standard deviations of the network are by comparing it to the residuals, i.e. the differences between the real and expected values. 

  \[ \chi^2 = \frac{1}{N} \sum_{i=1}^{N} \frac{(\theta_i - \mu_i)^2}{\sigma_i^2} \]

  \begin{itemize}
      \item {\bf $\chi ^2 \approx $ 1}: This indicates that the model fits the data well. The residuals are consistent with the expected variance, considering the uncertainties in the data. In other words, the model’s predictions differ from the observed values by approximately the expected amount, given the measurement uncertainties.
      
      \item {\bf $\chi ^2 < $ 1}: This suggests that the uncertainties in the data have been overestimated, as the deviations between the observed and predicted values are smaller than what would be anticipated given the uncertainties.
      
      \item {\bf $\chi ^2 > $ 1}:This indicates that the model does not fit the data well. The deviations between the model predictions and the observed data are larger than expected based on the uncertainties. This may suggest that the uncertainties in the data have been underestimated.
  \end{itemize}

  Reduced Chi-Squared is valuable for understanding the fit of the model relative to the uncertainties in the predictions.

\end{enumerate}
These metrics collectively offer a comprehensive assessment of the model's performance, highlighting both its accuracy and precision.

\section{Results} 
\label{sec:results}
In this section, we present the outcomes derived by performing inference of the IllustrisTNG model parameters using averaged galaxy cluster profiles.


\subsection{Inference with multiple cluster profiles} 

\begin{figure*}
 \centering
 \includegraphics[scale=0.20]{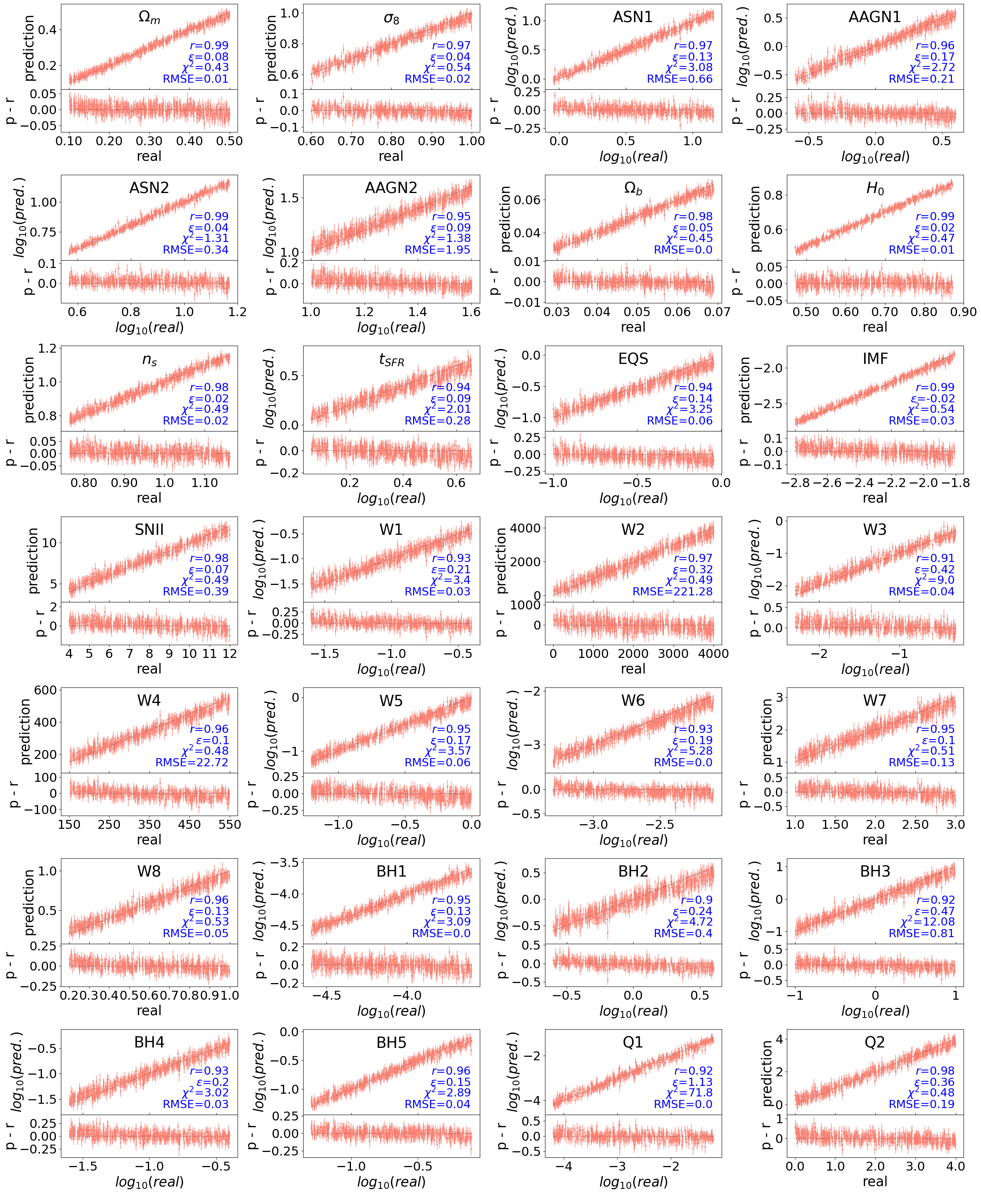}

     \caption{Inference results for the parameters in the IllustrisTNG model. The top panels for each parameter show the comparison between true values (x-axis) and inferred values (y-axis), with a black line representing the one-to-one correspondence. The bottom panels display the residuals (inferred minus true values) as a function of the true values to better illustrate the spread. 
     Each subplot corresponds to a different parameter, covering a wide range of cosmological and astrophysical parameters within the model. The correlation coefficient (\(r\)), root-mean-square error (RMSE), mean relative error, and reduced chi-squared (\(\chi^2\)) are reported within each top panel. The results demonstrate high accuracy and reasonable error margins across all parameters, as indicated by the strong correlation coefficients and relatively low residuals. }
\label{all_300000}
\end{figure*}

The primary aim of this study is to establish the theoretical framework for cosmological parameter inference using averaged or stacked galaxy cluster profiles by inferring cosmological and astrophysical parameters from simulated clusters in the IllustrisTNG model (parameters listed in Table \ref{tab:illustrisparams}). As a start, we establish a set of 300,000 points in the IllustrisTNG model parameter space, for which we emulate the five averaged profile quantities, at a constant mass of  $10^{14} M_{\odot}$. Of these points in parameter space, 70$\%$ are selected for training, while 15$\%$ are reserved for validation and testing, respectively. For each parameter space point, a 1-dimensional vector of concatenated averaged cluster profiles serves as the input to our network.

Once the model is trained and validated with five concatenated averaged profiles using 210,000 (70$\% \times$ 300,000) and 45,000 (25$\% \times$ 300,000) parameter space locations for training and validation, we test it using other independent 45,000 parameter space locations. The model returns for each input vector a posterior mean and standard deviation as prediction for each parameter of the IllustrisTNG model.  Fig. \ref{all_300000} shows the result of the prediction for each of the 45,000 individual points in parameter space belonging to the testing set. Figure \ref{all_300000} displays the prediction results for each of the 45,000 test points in parameter space, with the y-axis representing the network’s predicted values and the x-axis showing the actual ``target" or real values. The error bars indicate the uncertainty estimated by the network for each prediction.

We employ the metrics presented in Sec. \ref{subsec:3.4} to quantify the model performance, revealing high accuracy and reasonable error margins. The model notably excels in inferring cosmological parameters like $\Omega_{\rm m}$, $H_0$, and $\Omega_b$, each showing a correlation coefficient of 0.99, followed closely by $\sigma_8$ and $n_s$ with correlation coefficients of 0.98 and 0.97, respectively. All astrophysical parameters demonstrate correlation coefficients above 0.90, with some, such as the IMF slope and the ASN2, approaching 1.

To further understand the outstanding performance of our model, we examine the influence of individual parameters within the IllustrisTNG model by generating mean galaxy cluster profiles under various fiducial conditions in the IllustrisTNG parameter space, involving 27 fixed parameters and one variable parameter. This method allows for a thorough exploration of each parameter's impact and nuances on different profile types, also highlighting potential degeneracies. 

\begin{figure*}
 \centering
 \includegraphics[scale=0.25]{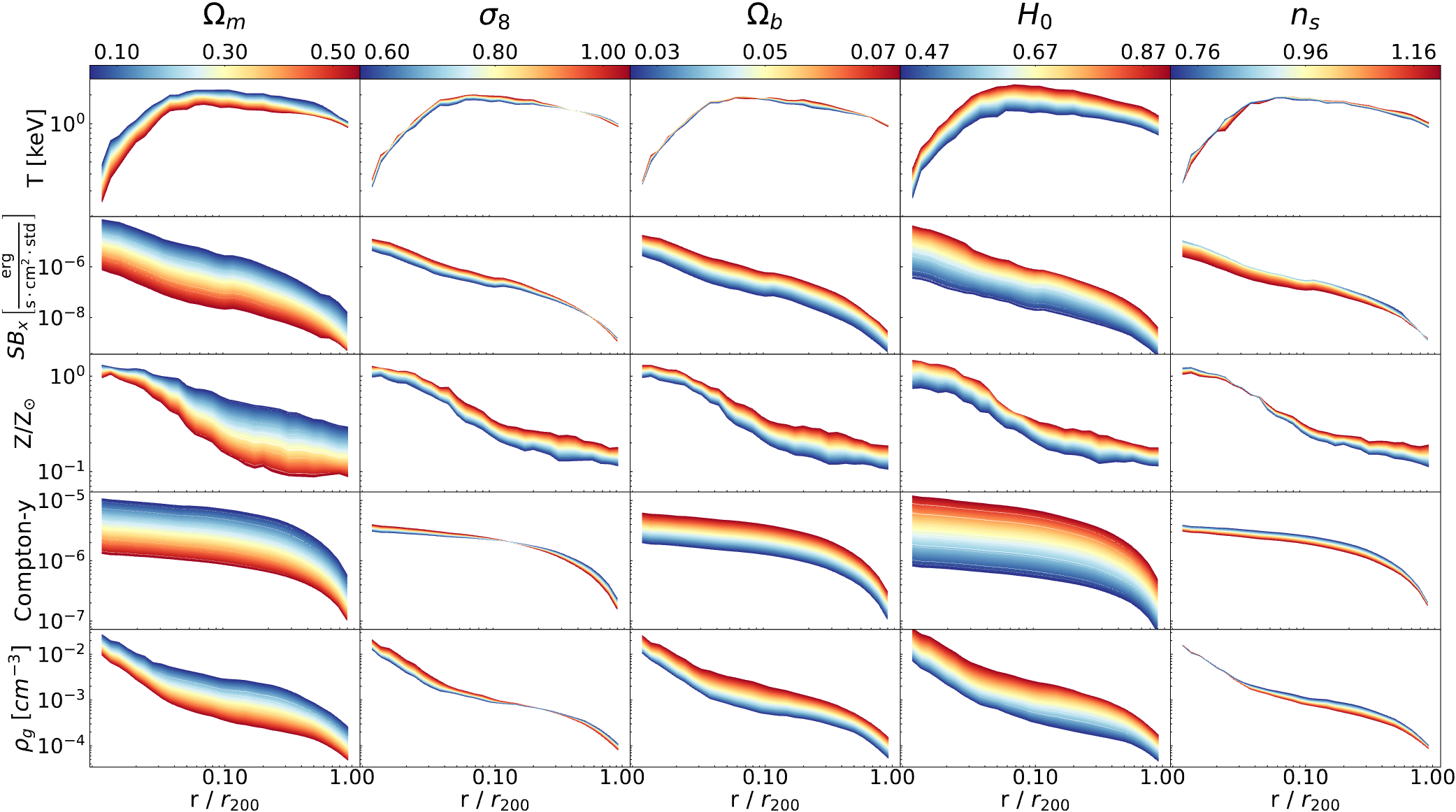}
     \caption{Mean galaxy cluster profiles illustrating the effects of varying each cosmological parameter individually. The rows correspond to different profiles: temperature (\(T\)), X-ray surface brightness (\(SB_X\)), normalized metallicity (\(Z/Z_{\odot}\)), Compton-y parameter, and gas density (\(\rho_g\)). Each column represents the impact of varying a single cosmological parameter (\(\Omega_{\rm m}\), \(\sigma_8\), \(\Omega_b\), \(H_0\), and \(n_s\)) across the full range of their values. The profiles are shown as a function of the radial distance normalized by \(R_{200}\). The color gradient within each panel highlights the spread 
     introduced into the profiles by the variations in the corresponding cosmological parameter, with the color scale representing the range of parameter values. This visualization demonstrates how changes in cosmological parameters affect different physical quantities in galaxy clusters, providing insight into the sensitivity of cluster profiles to cosmological variations.}
\label{profiles}
\end{figure*}

Fig. \ref{profiles} illustrates the impact of varying each cosmological parameter across its full range for each profile type. The results indicate that each parameter introduces distinct features at specific points in the profiles, which are not replicated by other parameters. This suggests that there are no significant degeneracies between the effects of different parameters on the profiles, as each produces unique, non-overlapping features. This holds true across all profile types.

Additionally, we observe that certain parameters cause greater variation than others. For example, \(\Omega_{\rm m}\) and \(H_0\) result in the most pronounced spread, followed by \(\Omega_b\), while \(\sigma_8\) and \(n_s\) show a more subtle effect on the profile variation. This pattern is reflected in the correlation coefficients shown in Fig. \ref{all_300000}, where \(\Omega_{\rm m}\) and \(H_0\) have the highest values, followed by \(\Omega_b\), with lower coefficients for \(\sigma_8\) and \(n_s\). This trend persists throughout the study, indicating that the former parameters are inferred more accurately than the latter.

Figs. \ref{all_profiles_r200_ngas_test2.pdf} - \ref{all_profiles_r200_xray_test2.pdf} in the appendix showcase the variations across all profile types for all 28 parameters, both astrophysical and cosmological. We can extend the argument above to this greater set of 28 parameters, as we observe that all of them introduce unique features. Even if we do not observe degeneracies among the parameters for any profile type, certain profile types seem to be more sensitive to specific parameters. For instance, astrophysical parameters like {\it IMF} have a mild effect on the SBx, gas density, and Compton-y profiles, but exert the strongest influence on the metallicity profiles.

Notably, while most profile types exhibit the greatest spread 
with variations in cosmological parameters, the metallicity profiles are equally, if not more, affected by astrophysical parameters. This is particularly evident with parameters such as the IMF,  ASN1, ASN2, and  W3. The distinct responses of these profiles to the underlying parameters suggest that their combined effects are key to optimal parameter inference. As demonstrated in Figs.  \ref{all_profiles_r200_ngas_test2.pdf} - \ref{all_profiles_r200_xray_test2.pdf}, there are no obvious degeneracies within the same profile type, indicating that the main challenge in parameter inference lies in unraveling the collective impact of multiple parameters—a task where machine learning techniques excel, thereby explaining the model's exceptional performance.

It is important to note, however, that we are working with stacked galaxy cluster profiles, which are noiseless and have a high number of bins (29 or 30, depending on the quantity). Additionally, the use of multiple stacked profiles helps reduce cosmic variance, contributing to the accuracy of our parameter inference. 

Finally, we note that although using 300,000 parameter space locations provides the most accurate results, we find that performance nearly saturates when using five profiles with 30,000 locations (see Sec. \ref{sec:A} in the appendix). As a result, in the following sections, we will proceed with 30,000 parameter space locations when working with all five profile types, and reserve the use of 300,000 locations for cases involving only a single profile type. This balance between accuracy and computational efficiency ensures a robust parameter inference while optimizing the use of resources.

\subsection{Inference with individual profile types}

To discern the unique contributions of each profile type, we conducted independent analyses using 300,000 profiles per type while employing the same architectural framework. The upper left panel of Fig. \ref{corr_coefficient_unlog} displays the correlation coefficients for each parameter inferred using different profile types.

Generally, some parameters, such as the cosmological parameters $\Omega_{\rm m}$,  $H_0$, and $n_s$ and the astrophysical parameters ASN1, ASN2, and IMF show higher correlation coefficients across most profile types, indicating more robust inference. On the other hand, most of the wind parameters (W1-W8) and black hole (BH1-BH5) parameters tend to have lower correlation coefficients, suggesting they are more challenging to infer accurately.

Each profile type is sensitive to different model parameters and with a different strength:

\begin{itemize}

\item {\bf Gas density profiles:} These profiles show high correlation coefficients, particularly for cosmological parameters such as $\Omega_{\rm m}$ and $\Omega_b$. These parameters represent the density ratios of matter and baryonic matter, respectively, to the critical density of the universe. Their strong correlations with gas density suggest that these profiles are still sensitive, even if astrophysical phenomena are present, to small variations in the overall mass and baryonic content in clusters, which are directly influenced by these cosmological densities. A similar rationale applies to $\sigma_8$, which describes the amplitude of matter fluctuations and is best inferred by the gas density profile which outperforms the rest of the quantities. The density profile also shows high correlation coefficients for wind parameters, black hole, and black hole feedback parameters. In the case of the wind parameters, these parameters govern aspects of wind energy, momentum, and travel distance (W1-W4) as well as energy reduction and wind particle decoupling (W5-W8). In the case of the BH parameters, they govern the initial BH mass, accretion rates, and the efficiency in energy conversion (BH1-BH5) as well as the energy injection strength and direction of the inhomogeneous feedback mode (Q1-Q2). Thus, all these parameters regulate the distribution of gas inside the cluster, leaving a distinctive imprint on the gas density profiles. 

\item {\bf Temperature profiles:} The temperature profile performs well across most parameters, particularly cosmological and some feedback parameters. For instance, it outperforms all other quantities in most wind parameters. Galactic winds, driven by supernovae, inject energy into the interstellar and intracluster medium, heating the gas and altering its temperature. Temperature is a sensitive indicator of feedback processes, which are critical in regulating star formation and redistributing gas within clusters. Wind-driven feedback mechanisms inject thermal energy into the surrounding gas, raising its temperature. The temperature profile is sensitive to this, reflecting the cumulative effect of these winds over time and space within the cluster. As a result, the temperature profile becomes a key observable in understanding how winds modulate the thermal state of the gas.

\item {\bf Metallicity profiles:} The metallicity profiles show weaker performance for cosmological parameters like $\sigma_8$, and $H_0$. This is likely because cosmological parameters govern large-scale structure and the overall mass distribution of the universe, which influence the gas distribution and temperature more directly than metallicity. Metallicity is more of a tracer of the cumulative effects of star formation and feedback processes over time, making it less sensitive to the direct effects of cosmological parameters. Thus, we observe that the metallicity profile tends to perform relatively better for parameters associated with stellar feedback processes (e.g., ASN1, ASN2) and star formation (e.g., IMF). This pattern can be attributed to the fact the stellar content and the corresponding feedback in the form of supernovae can enrich the intracluster medium with metals, thereby altering the metallicity distribution. Since these processes directly influence the metallicity, the metallicity profile is more sensitive to changes in these parameters, leading to higher correlation coefficients. 

Although winds influence metal distribution, their effect on temperature is more pronounced than on metallicity. Metals are less efficiently transported by winds than thermal energy. Due to their higher atomic weight and radiative properties, metals cool more quickly and mix with the intracluster medium (ICM) in a slower and more localized fashion. This explains why changes in metallicity are more spatially confined than changes in temperature, making metallicity profiles less sensitive to wind parameters.

We also observe that the metallicity profiles seem to be less sensitive to AGN feedback and BH parameters. The metallicity reflects the long-term history of star formation and supernova activity, gradually enriching the ICM with metals. This process occurs over longer timescales compared to the more episodic and intense events associated with AGN feedback. While AGN feedback can have significant short-term impacts on the local environment, such as heating and gas displacement, these effects do not necessarily result in immediate changes in metallicity. The metals might have already been distributed by earlier generations of stars and supernovae, with AGN feedback playing a role more in redistributing or removing gas rather than enriching it.

\begin{figure*}
 \centering
 \includegraphics[scale=0.34]{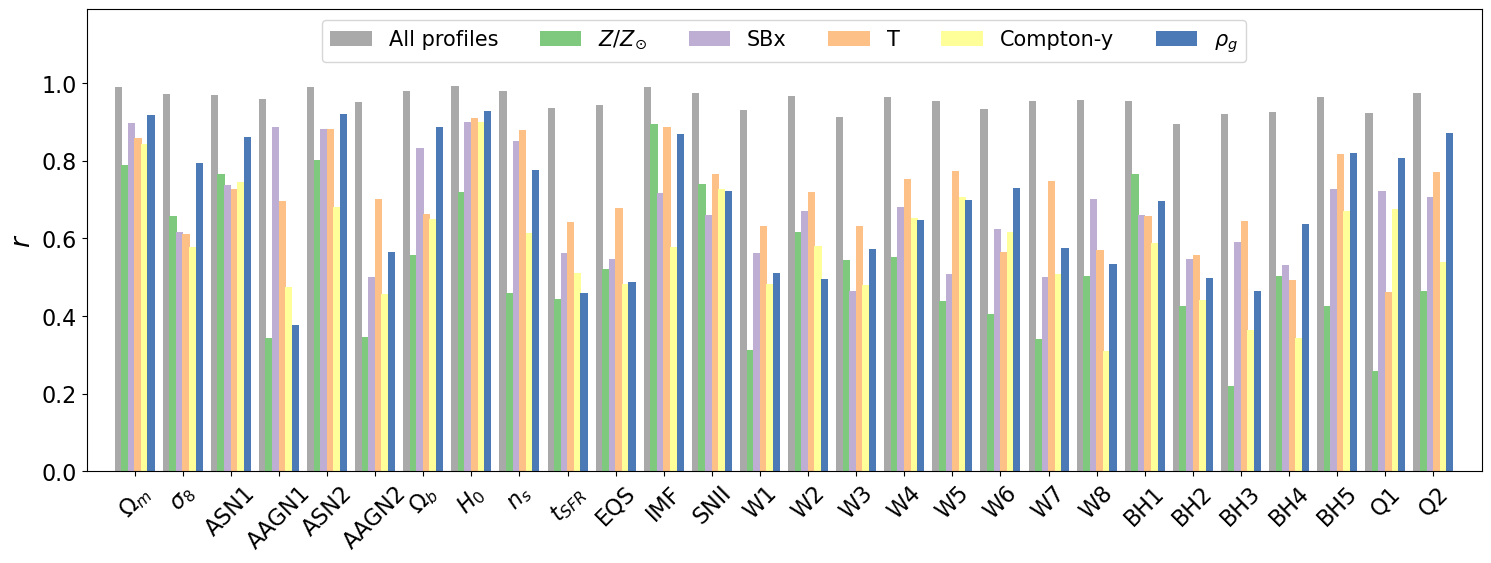}
 
     \caption{Correlation coefficient (\(r\)) values for the inference of all IllustrisTNG parameters, analyzed separately for different types of mean galaxy cluster profiles. The different colors represent the various types of profiles used: gas density (\(\rho_g\)), temperature (T), X-ray surface brightness (SBx), metallicity (Z), and Compton-y parameter (Y). Each set of bars corresponds to a different IllustrisTNG parameter, listed along the x-axis, with the correlation coefficient on the y-axis. The study utilized 300,000 profiles for each type, with 70$\%$ allocated for training and the remaining 30$\%$ used for validation and testing. The plot illustrates how the choice of profile type affects the accuracy of parameter inference, with varying correlation coefficients observed across the different parameters and profile types. }
\label{corr_coefficient_unlog}
\end{figure*}

\item {\bf X-Ray Surface Brightness profiles:} The surface brightness profiles generally show intermediate correlation coefficients for most parameters. It tends to perform better than Compton-y but often underperforms compared to gas density and temperature profiles. This suggests that while surface brightness is a useful observable, it may not be sensitive to the full extent of the physical processes influencing certain parameters as effectively as other profiles.

The surface brightness profile tends to perform relatively well for cosmological parameters such as $\Omega_{\rm m}$, $\sigma_8$, and $\Omega_b$. Since SBx is a measure of the X-ray emission, it is sensitive to the amount and distribution of gas in clusters, which in turn is influenced by the large-scale distribution of matter and the potential well of the clusters which is governed by cosmological parameters. Regions of higher matter density will have more gas, higher temperatures, and thus stronger X-ray emission. This makes the SBx profile a good tracer of cosmological parameters, especially those that affect the large-scale structure of the universe, such as $\Omega_{\rm m}$ and $\sigma_8$. However, because SBx is an integrated quantity along the line of sight, it averages the X-ray emission over the entire cluster. This can smooth out local variations in gas density and temperature, especially those caused by astrophysical processes like feedback from AGNe or supernovae. As a result, while SBx is sensitive to the broad effects of cosmological parameters (which impact the overall matter distribution and cluster formation), it may not be as sensitive to localized phenomena like feedback, which affect temperature and gas density more directly.

\item {\bf Compton-y profiles:} The Compton-y profiles tend to show lower correlation coefficients across most parameters compared to other profiles like gas density and temperature. This suggests that the Compton-y profile is less effective at capturing the nuances of many parameters in the IllustrisTNG model, particularly those related to feedback and small-scale processes.

Despite its generally lower performance, the Compton-y profile does exhibit good correlation for parameters that directly influence the large-scale thermal properties of the gas, such as certain cosmological parameters ($\Omega_{\rm m}$, $\sigma_8$) and feedback parameters that affect the overall thermal energy content of the ICM. 

The Compton-y profile generally underperforms for feedback-related parameters (e.g., ASN1, ASN2, AAGN1, AAGN2) and wind parameters (W1-W8). This could be because these parameters often involve localized or dynamic processes that impact the temperature and density of the gas in ways that may not significantly alter the overall thermal pressure integrated along the line of sight. Consequently, these effects are not well captured by the Compton-y profile, which averages over these localized variations.

Nevertheless, although the Compton-y profile appears less effective at capturing detailed, localized processes, its smoother nature may make it less sensitive to the specific subgrid models used in the simulation. As a result, it could still be very valuable when transitioning this method to future observational applications.

\end{itemize}

\subsection{Different mass bins}

\begin{figure*}
 \centering
 \includegraphics[scale=0.36]{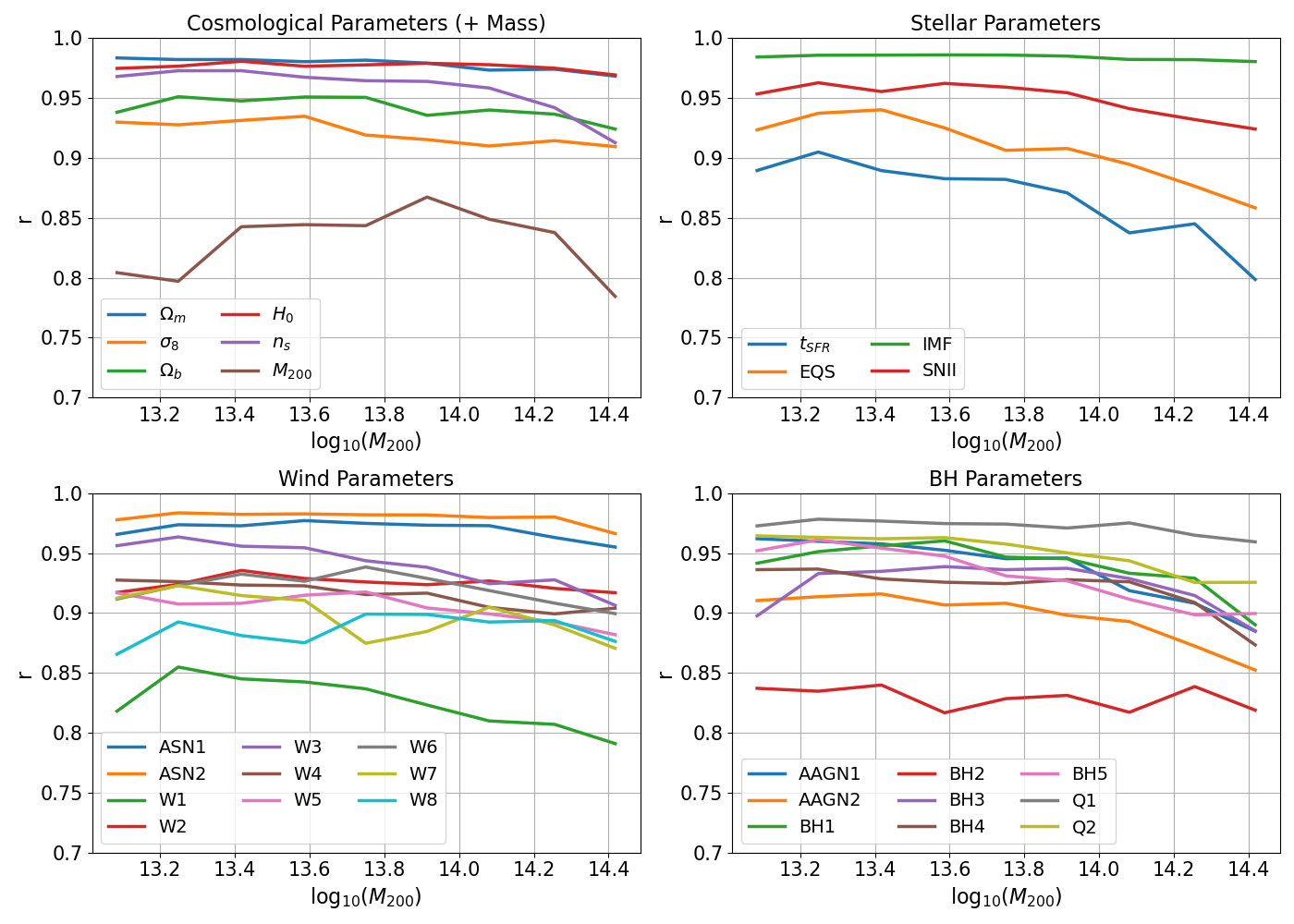}

     \caption{Correlation coefficients (\(r\)) for the inference of various parameters as a function of galaxy group/cluster mass. Upper left panel: Correlation coefficients for cosmological parameters and mass. Upper right panel: Correlation coefficients for star formation and feedback parameters. Lower left panel: Correlation coefficients for wind parameters. Lower right panel: Correlation coefficients for black hole (BH) and active galactic nucleus (AGN) parameters. The x-axis represents the mass of the galaxy groups or clusters, while the y-axis shows the correlation coefficient values. Each line represents a different parameter within each category. The minimum value of the lowest correlation coefficient is approximately 0.8 across all mass bins, indicating that the inference method is robust for galaxy groups and clusters across a mass range of \(10^{13}\) to \(3 \times 10^{14}\) solar masses. This robustness suggests reliable parameter inference across the entire mass range studied. }
\label{corr_mass_r200}
\end{figure*}

Our previous sections focused on mean cluster profiles for galaxy clusters with a mass of \(1 \times 10^{14} M_{\odot}\). Here, we expand our analysis to explore how predictions vary across different galaxy cluster masses. 

To do this, we generated new sets of mean galaxy cluster profiles in 30,000 different parameter space locations, where we treated the mass of the cluster as an extra parameter so that our model has now 29 parameters (28 IllustrisTNG parameters + mass). We varied the value of the mass to cover the full spectrum of galaxy group and cluster masses in our simulations, this is from $1 \times 10^{13} M_{\odot}$ to $3 \times 10^{14} M_{\odot}$. 
The neural network’s performance in parameter inference across these different masses is illustrated in Fig. \ref{corr_mass_r200}.

In the upper left panel of Fig. \ref{corr_mass_r200}, the correlation coefficients for cosmological parameters (\(\Omega_{\rm m}\), \(\sigma_8\), \(\Omega_b\), \(H_0\), \(n_s\)) and mass are consistently high across all mass bins. This indicates that the inference of these parameters is robust across various galaxy group and cluster masses, with minimal variation in accuracy as mass increases. Although the correlation for the mass parameter shows a slight dip, it remains above 0.8 overall, indicating stable performance.

In contrast, the upper right panel of Fig. \ref{corr_mass_r200} shows that the correlation coefficients for stellar parameters tend to decrease with increasing mass, particularly for parameters like the Maximum Star Formation Timescale (\(t_{SFR}\)) and the Factor for Softer EQS. This trend suggests that stellar feedback processes, as captured by these parameters, become more challenging to infer accurately in more massive clusters.

A similar pattern is observed for the wind parameters in the lower left panel and for the BH and AGN feedback parameters in the lower right panel, where most parameters show a slight decrease in correlation coefficients with increasing mass. 

This decrease in inference accuracy with increasing mass can be attributed to the fact that, as clusters grow more massive, internal processes such as star formation, supernova feedback, and AGN activity have a relatively smaller impact on overall cluster properties. Instead, gravitational dynamics and large-scale structure become the dominant factors driving massive structures closer to self-similarity. As a result, the accuracy of inferring these parameters may decrease with mass, as cluster properties become more governed by simpler, mass-dependent scaling laws rather than detailed processes.

Nevertheless, despite the observed decreases in correlation coefficients with mass, the minimum value of the lowest correlation coefficient is approximately 0.8 across all mass bins. This indicates that the inference method remains generally robust across a wide range of galaxy groups and cluster masses, from \(10^{13}\) to \(3 \times 10^{14}\) solar masses. This robustness suggests that, even though certain parameters are more sensitive to mass and exhibit varying performance, mean galaxy cluster profiles still contain key information about both the processes occurring within these systems and the cosmology of the universe they inhabit, across the entire mass range studied.

\subsection{Noisy profiles}

\begin{figure}
 \centering
 \includegraphics[scale=0.25]{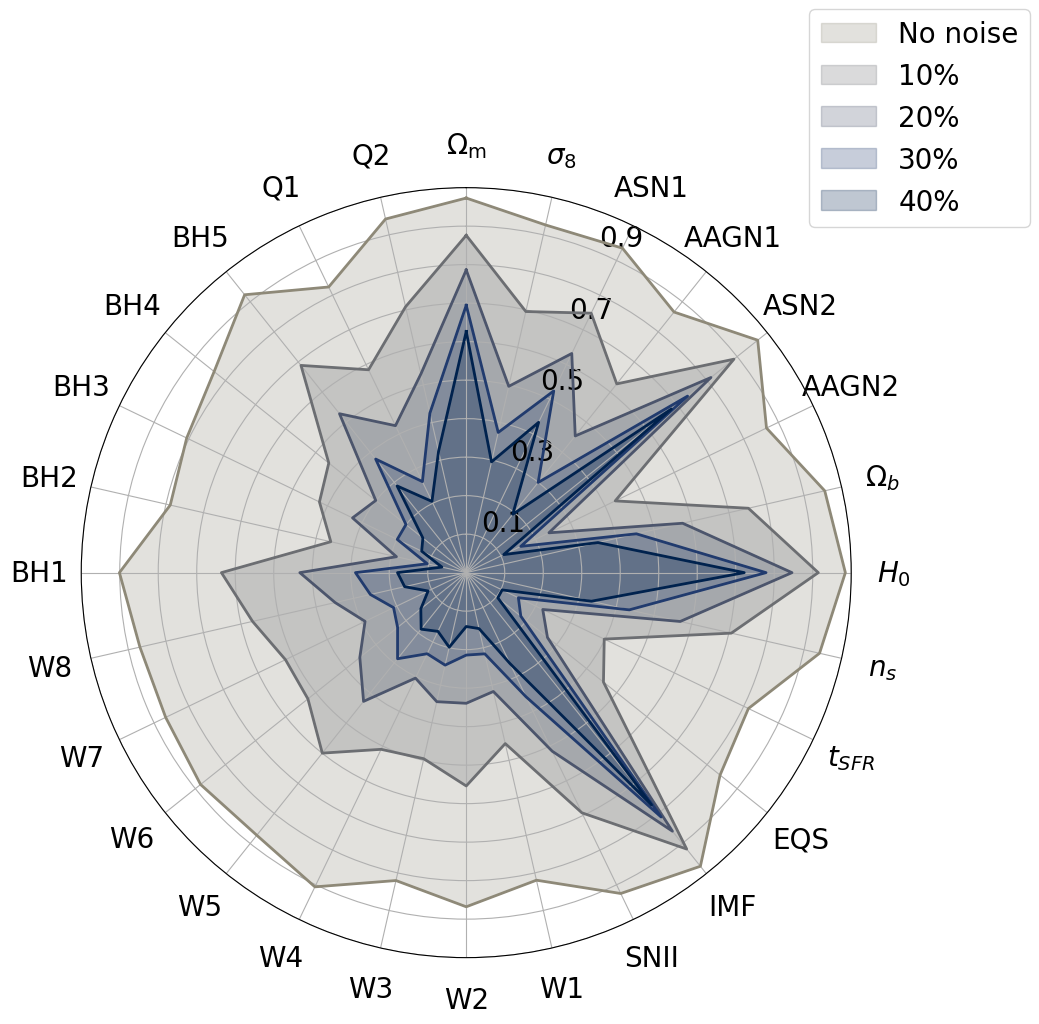}
     \caption{Radar chart displaying the correlation coefficients for various IllustrisTNG parameters under different levels of Gaussian noise. The radial distance from the center represents the correlation coefficient, with higher values indicating stronger correlations. Each parameter is represented by a specific symbol on the perimeter of the chart, corresponding to the different physical or cosmological parameters within the IllustrisTNG model. The different shaded regions correspond to varying noise levels, ranging from no noise to 40$\%$ Gaussian noise. As noise increases, the correlation coefficients generally decrease, indicating a reduction in the accuracy of parameter inference. The chart visually demonstrates how increasing noise impacts the reliability of inferred parameters, with the outermost regions (no noise) showing the highest correlation, and the innermost regions (40$\%$ noise) showing the lowest correlation. }
\label{corr_coef_unlog_noise_rad}
\end{figure}

Galaxy cluster profiles observed in astronomical data are often impacted by various sources of noise, typically quantified by the signal-to-noise ratio (S/N). It is common to perform spectral modeling in each spatial pixel to ensure a minimum S/N, usually having a value of 10 to 100 \citep{2010A&ARv..18..127B}, which is then processed to produce either a projected or deprojected profile. To evaluate how noise affects our parameter inference, we simulate the expected observational noise by introducing Gaussian noise into our profiles. For each bin, we added a specific percentage of random Gaussian noise, creating datasets with noise levels of 10$\%$, 20$\%$, 30$\%$, and 40$\%$, corresponding to S/N ratios of 10, 5, 3.3, and 2.5, respectively. This approach allows us to study the limit of S/N = 10, above which we anticipate improved results, while also examining how inference accuracy degrades with increasingly noisy data. These modified profiles were then used for the training, validation, and testing of our model.

The effect of the introduced noise in the parameter inference is depicted in Fig. \ref{corr_coef_unlog_noise_rad}. As Gaussian noise increases from 0$\%$ to 40$\%$, there is a general decline in the correlation coefficients for all parameters, indicating that noise consistently reduces the accuracy of parameter inference. However, the extent of this decline differs between the parameters. For example, parameters like \(\Omega_{\rm m}\), \(\Omega_b\), and \(H_0\) maintain relatively high correlations even at 40$\%$ noise, suggesting that these cosmological parameters are more robust to noise. Similarly, certain astrophysical parameters, such as ASN2 and IMF, also exhibit high correlation levels despite the added noise. 

In contrast, other parameters, such as W7 among the wind parameters, and Q1 and BH2 among the black hole parameters, experience a sharper decline in correlation coefficients, indicating greater sensitivity to noise. This suggests that the inference of these parameters is more vulnerable to the degradation of input data quality.

Parameters associated with similar physical processes or feedback mechanisms tend to exhibit similar patterns of sensitivity to noise. For example, many wind parameters (W1-W8) and black hole feedback parameters (BH1-BH5) show a clustered response to increasing noise, with significant declines in correlation, suggesting that certain processes are uniformly more challenging to infer accurately in the presence of noise.

The relative ranking of parameters by their correlation coefficients remains fairly stable across different noise levels. Parameters that start with high correlations in the no-noise scenario (e.g., \(\Omega_{\rm m}\), \(n_s\)) generally maintain higher correlations even as noise increases, compared to those that begin with lower correlations. This consistency implies that while noise affects all parameters, those that are inherently more robust—likely due to their strong influence on large-scale cluster properties—remain more reliable even under noisy conditions.

This analysis suggests that when noise is introduced, the results are driven less by the non-degeneracy of parameters and more by the strength of each parameter's influence on the profiles. Some parameters may only affect the profiles in specific ranges where the data is highly correlated. If this correlation is disrupted by noise, the network's performance declines. However, certain cosmological parameters like \(\Omega_{\rm m}\), \(\Omega_b\), and \(H_0\), as well as astrophysical parameters such as IMF and ASN2, have an impact on the profiles that remains robust to noise. We note that these parameters are also the ones that can be inferred well with lower training data as shown in Fig. \ref{corr_coef_unlog_base_rad} of appendix \ref{sec:A}. These parameters generate significant variation in the profiles (see Figs. \ref{all_profiles_r200_ngas_test2.pdf} - \ref{all_profiles_r200_xray_test2.pdf}), and the strength of this correlation allows them to be inferred accurately even under extreme noise conditions.

\subsection{Inference accuracy and radial cut}

\begin{figure*}
 \centering
 \includegraphics[scale=0.34]{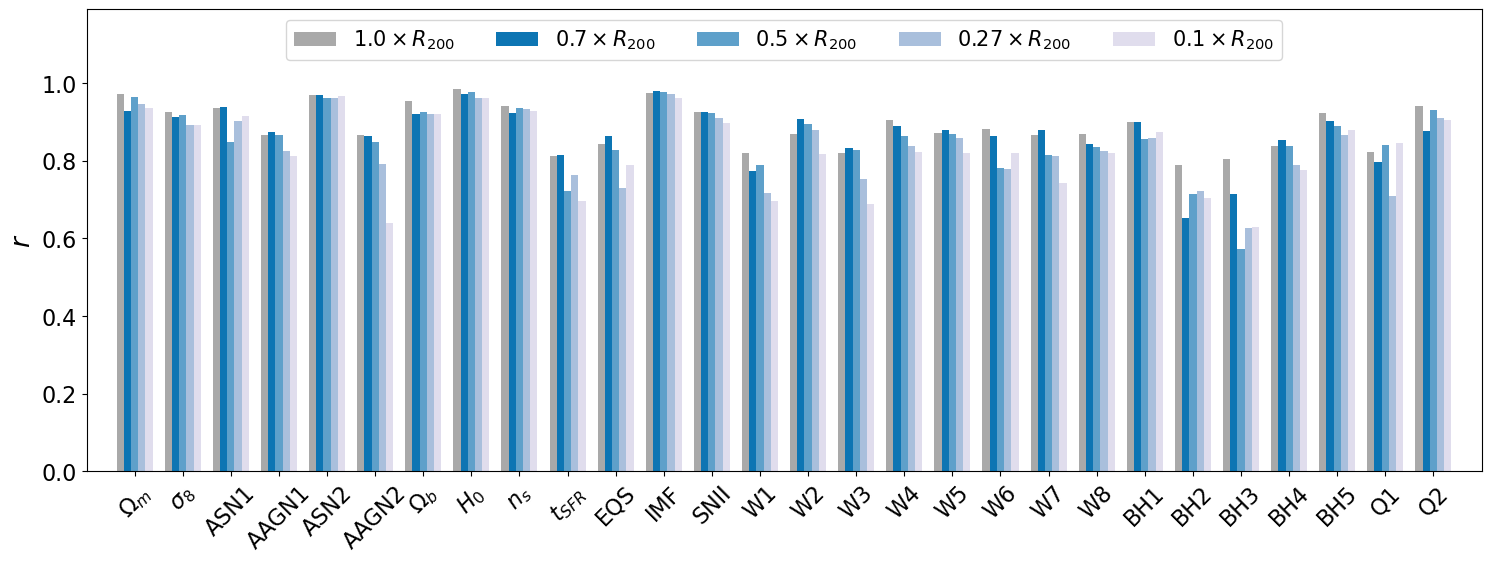}
     \caption{Bar chart displaying the correlation coefficients (\(r\)) for various IllustrisTNG parameters as a function of different radial cuts within the cluster, ranging from \(1.0 \times R_{200}\) to \(0.1 \times R_{200}\).  The parameters are listed along the x-axis, with each set of bars representing a different parameter, while the color gradient indicates the radial cut used for the analysis. The results show that while correlation coefficients generally decrease with smaller radial cuts, the overall change is not very large. This suggests that a significant portion of the information required for accurate parameter inference is concentrated within the innermost \(0.1 \times R_{200}\) of the cluster's radius.}
\label{corr_coef_unlog_radcut}
\end{figure*}

\begin{figure*}
 \centering
 \includegraphics[scale=0.344]{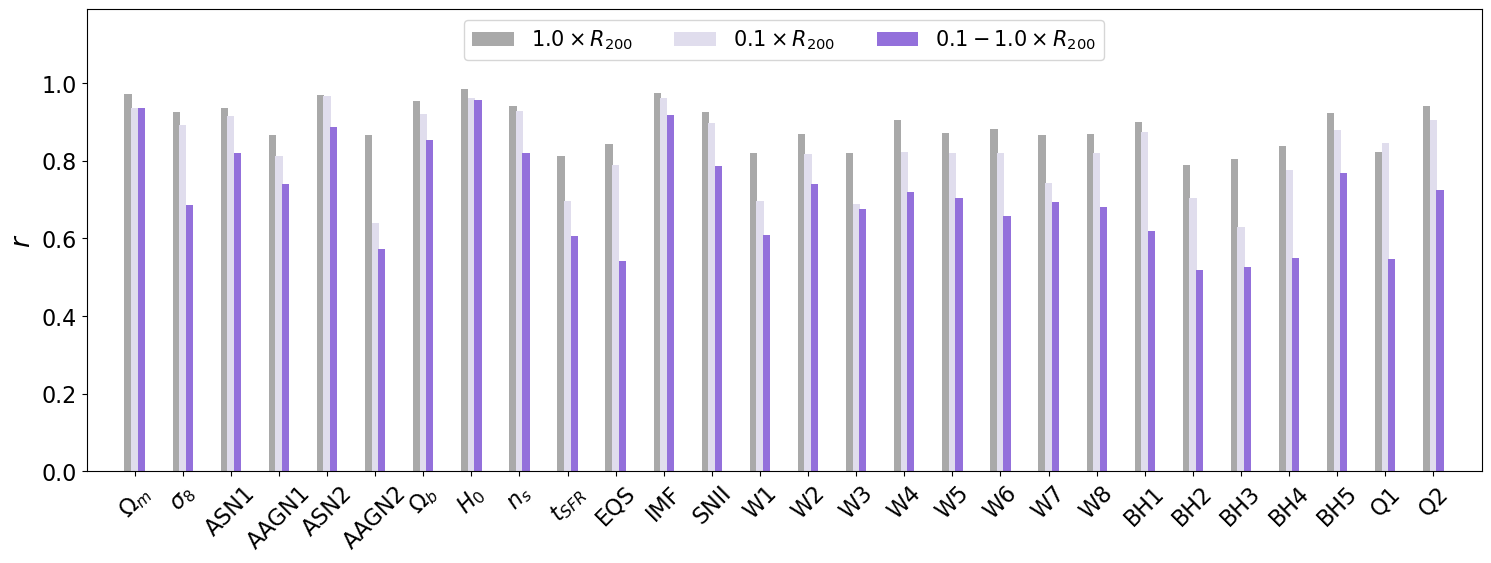}
\includegraphics[scale=0.344]{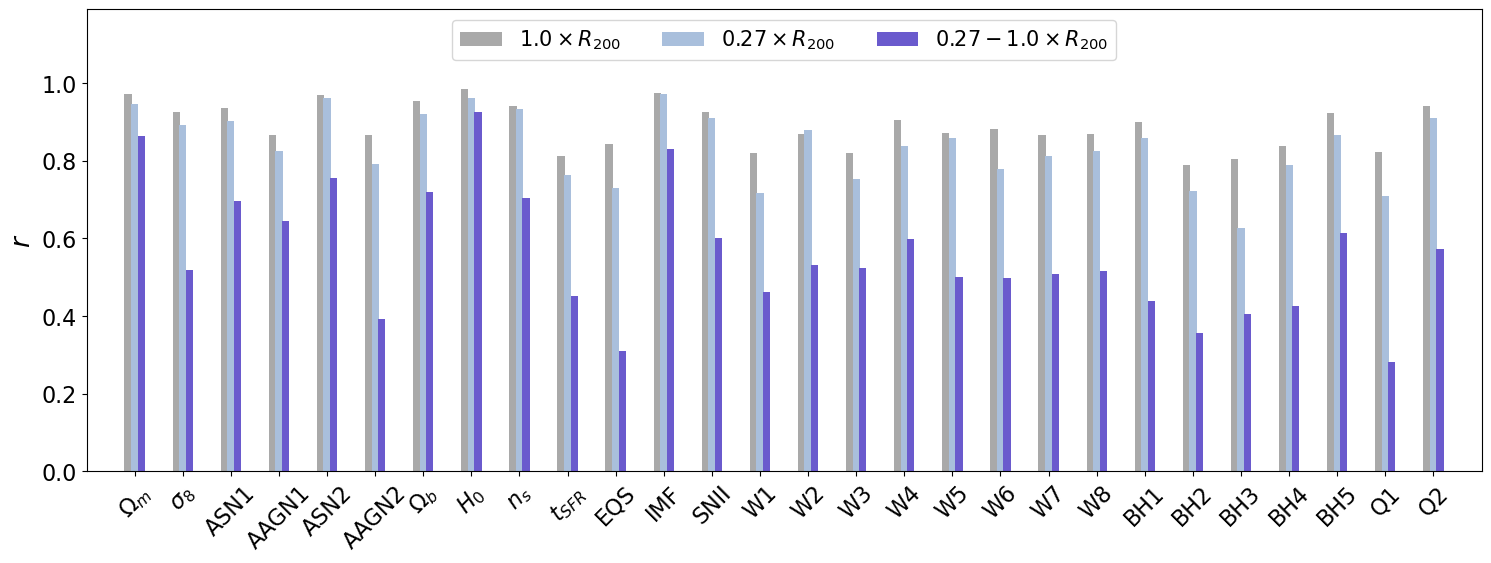}
     \caption{Bar charts comparing the correlation coefficients (\(r\)) for various IllustrisTNG parameters across different radial ranges within the cluster. The left panel shows the correlation coefficients for the full radial range (\(1.0 \times R_{200}\), gray), the inner region (\(0.1 \times R_{200}\), light purple), and the outer shell between \(0.1 \times R_{200}\) and \(1.0 \times R_{200}\) (dark purple). The right panel displays the correlation coefficients for the full radial range (\(1.0 \times R_{200}\), gray), the central region (\(0.27 \times R_{200}\), light blue), and the outer shell between \(0.27 \times R_{200}\) and \(1.0 \times R_{200}\) (dark blue). In both panels, each set of bars represents a different parameter, with the correlation coefficient on the y-axis. The results indicate that the central regions (up to \(0.1 \times R_{200}\) or \(0.27 \times R_{200}\)) generally provide the highest correlation coefficients, suggesting they contain the most critical information for accurate parameter inference. The outer shells show a significant drop in correlation, especially in the right panel, highlighting the diminished contribution of the outer regions to the overall accuracy. These findings underscore the importance of the inner cluster regions in the inference of the model parameters.}
\label{corr_coef_unlog_extract}
\end{figure*}

\begin{figure*}
 \centering
 \includegraphics[scale=0.16]{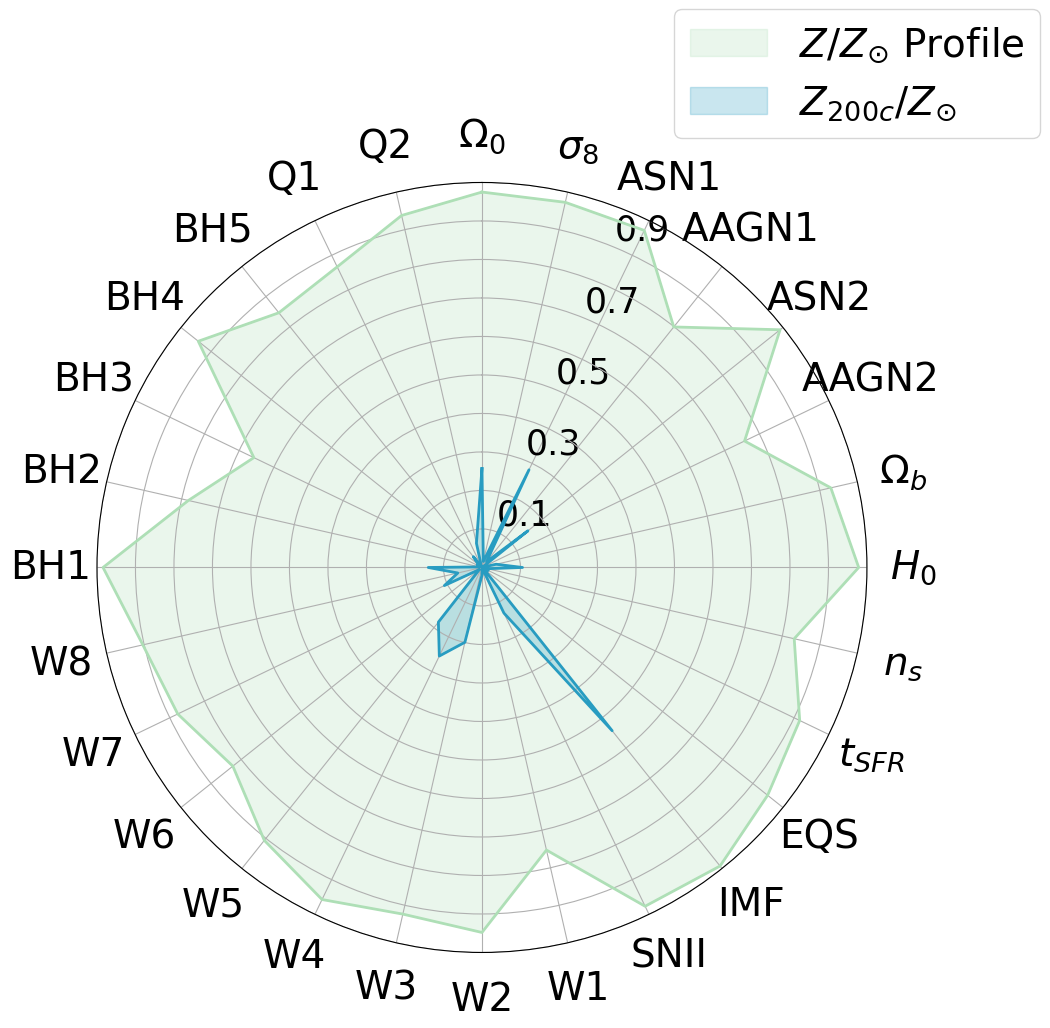}
\includegraphics[scale=0.16]{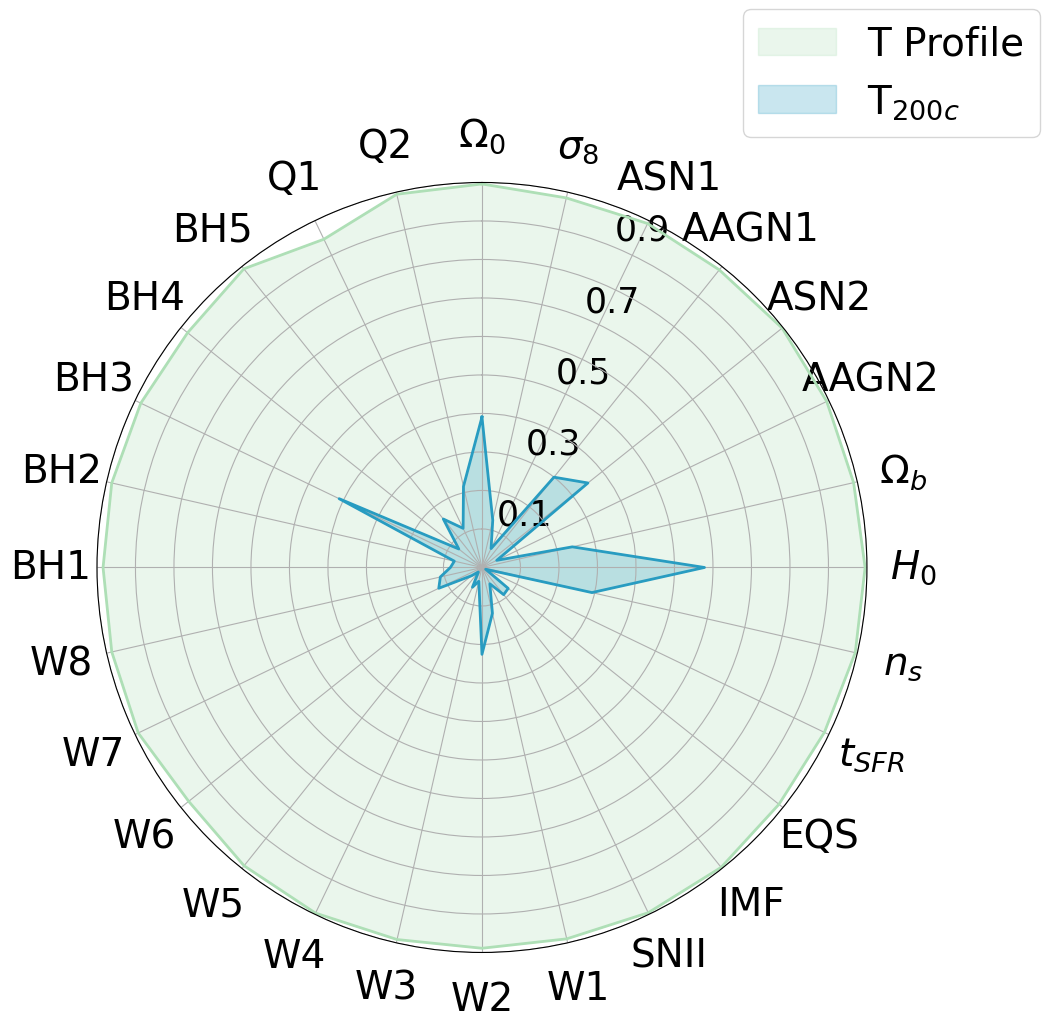}
\includegraphics[scale=0.16]{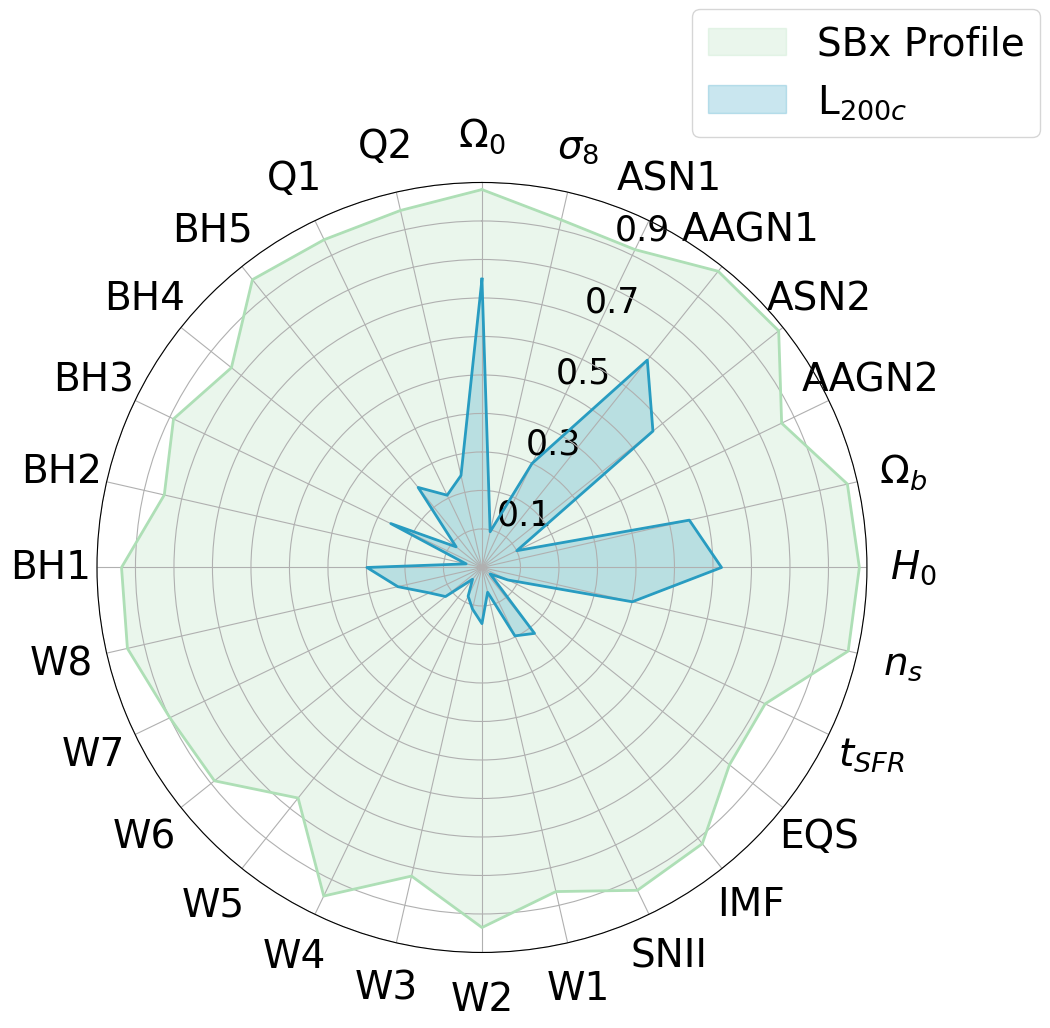}
\includegraphics[scale=0.16]{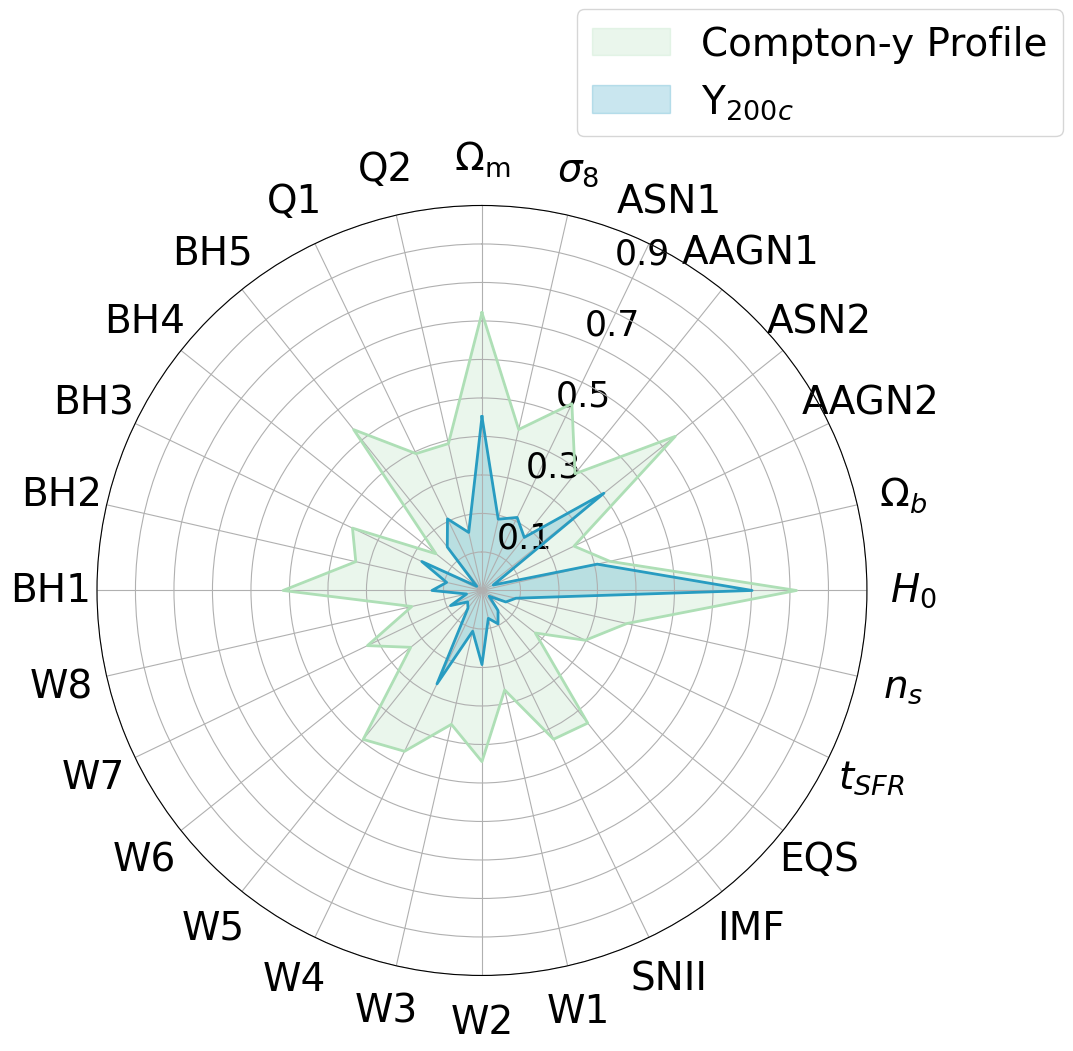}
\includegraphics[scale=0.16]{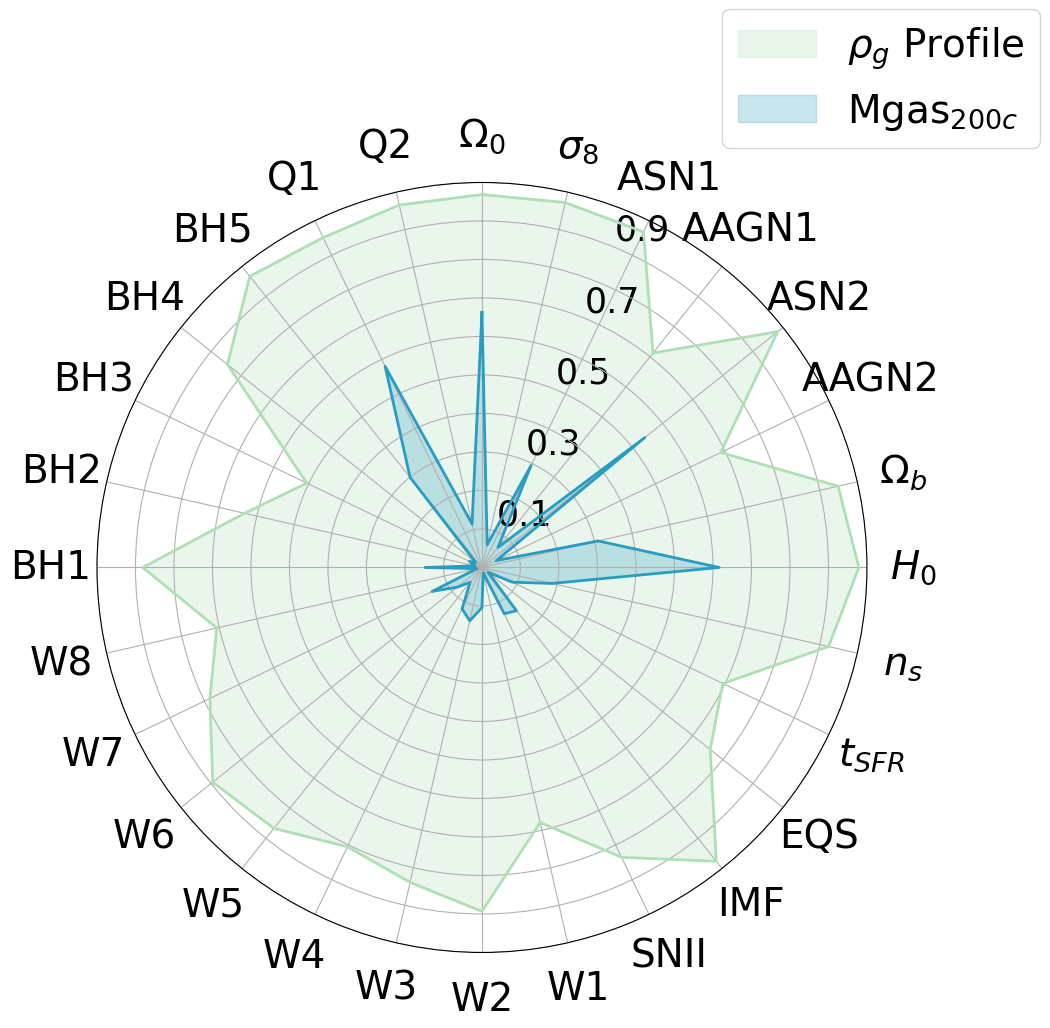}
     \caption{Radar charts showing the correlation coefficients for various IllustrisTNG parameters, with the radial axis representing the correlation coefficient values. Each chart compares the results obtained from using the profile of a specific quantity (light-shaded region) against those obtained from the integrated quantity up to \(R_{200}\) (dark-shaded region). The quantities examined are (top left) metallicity (\(Z/Z_{\odot}\)), (top center) temperature (\(T\)), (top right) surface brightness (S\(_{X}\)), (bottom left) Compton-y parameter, and (bottom right) gas density (\(\rho_g\)). The light-shaded regions generally encompass larger areas, indicating that using profiles provides higher correlation coefficients and, consequently, more accurate parameter inference compared to using integrated quantities.}
\label{corr_coef_unlog_integrated}
\end{figure*}

To determine which regions within galaxy clusters provide the most significant information for parameter inference, we performed an analysis by progressively excluding the outer parts of the cluster. Specifically, we applied radial cuts at 70\%, 50\%, 27\%, and 10\% of the virial radius, thereby removing the outer regions beyond these thresholds. Fig. \ref{corr_coef_unlog_radcut} shows the correlation coefficients obtained for each of these radial cuts compared to using the full virial radius. The correlation coefficients (\(r\)) for various IllustrisTNG parameters are plotted as a function of different radial cuts within the cluster, ranging from 1.0 \(R_{200}\) (i.e., the entire virial radius) down to 0.1 \(R_{200}\). Each set of bars represents a different parameter, with the color gradient indicating the specific radial cut used in the analysis. 

Figure \ref{corr_coef_unlog_radcut} shows that as the radial cut decreases (from 1.0 \(R_{200}\) to 0.1 \(R_{200}\)), the correlation coefficients generally decrease for most parameters. However, this decline is modest, suggesting that while the outer regions do contribute to parameter inference, the bulk of the relevant information is concentrated in the inner region, particularly within the innermost 0.1 \(R_{200}\). Interestingly, for some parameters, shorter profiles yield slightly better results, which is counterintuitive since reducing the number of bins should reduce the information available to the network. We attribute this to the epistemic uncertainty in the network, arising from the inherent limitations of any model, and improvements to the architecture could potentially reduce this uncertainty. These results suggest that although parameter inference accuracy declines slightly as the radial cut decreases, strong results can still be achieved for most parameters, even when the radius is truncated to 10$\%$.

The upper panel of Fig. \ref{corr_coef_unlog_extract} compares the correlation coefficients (\(r\)) for various IllustrisTNG parameters when using only the innermost 0.1 \(R_{200}\) region (light purple) versus the combined outer region from 0.1 to 1.0 \(R_{200}\) (dark purple). Each set of bars corresponds to a different parameter, with the correlation coefficient displayed on the y-axis. The results demonstrate that the innermost 0.1 \(R_{200}\) region generally yields higher correlation coefficients across most parameters, indicating that this central region contains the most critical information for accurate parameter inference. The outer region contributes less to the overall correlation, as reflected by the lower coefficients when this region is used in isolation.

The lower right plot offers a similar comparison, this time between the central 0.27 \(R_{200}\) region (light blue) and the outer shell from 0.27 to 1.0 \(R_{200}\) (dark blue). Similar to the previous comparison, the central 0.27 \(R_{200}\) region shows higher correlation coefficients than the outer shell. The outer shell exhibits a more pronounced drop in correlation coefficients compared to the central region, highlighting that while it adds some information, its contribution to overall accuracy is less significant. This reinforces the conclusion that the cores of galaxy clusters are crucial for accurate parameter determination, which is not entirely unexpected since cluster cores tend to be more affected by astrophysical processes such as cooling, strong star formation bursts, and the effects of AGN feedback.

\subsection{Integrated quantities} 

Previous studies have focused on inferring some of the IllustrisTNG parameters using the integrated \(Y_{200c}\) \citep{2024ApJ...968...11L}. In this work, we extend the analysis to evaluate the inference capabilities of all integrated quantities and compare them to the capabilities derived from using the full cluster profile. To achieve this, we generate average galaxy cluster profiles in a specific parameter space location for 30 different masses, uniformly spaced in log space from \(10^{13}\) to \(3 \times 10^{14}\) solar masses. We select a total of 300,000 of these parameter space locations and create 30 averaged galaxy cluster profiles for each of them. 

Our input vector, representing a single location in parameter space, consists of one profile per mass bin, with each profile containing either 30 or 29 bins, depending on the quantity, resulting in a total of 900 or 870 values. For the integrated quantities, all profiles are collapsed into a single value, so each input vector will contain 30 values, one for each mass bin.

Figure \ref{corr_coef_unlog_integrated} demonstrates the correlation coefficients for the IllustrisTNG parameters, comparing the performance of using full galaxy cluster profiles (light shaded regions) versus integrated quantities up to \(R_{200}\) (dark shaded regions) for each individual quantity. Across all five radar charts, it is evident that using the full profiles generally results in higher correlation coefficients, implying more accurate parameter inference compared to the use of integrated quantities. The size of the light-shaded regions relative to the dark-shaded ones provides a visual representation of this difference.

The extent to which the profile outperforms the integrated quantity varies by parameter. Temperature profiles stand out most, as profiles achieve correlation coefficients above 0.9 for all parameters, while the integrated quantity does not surpass 0.6. This suggests that the detailed temperature profile carries a wealth of information that is lost when collapsing the profile into a single integrated quantity, highlighting the importance of spatial variations in accurately constraining model parameters. These results are followed by the gas density, metallicity, and surface brightness, which also show a better correlation coefficient for the profiles as for the integrated quantities.

For the Compton-y parameter, the difference between full profiles and integrated quantities is less pronounced. While full profiles still offer a slight advantage, the performance gap is narrower. This can be explained by the smoother nature of Compton-y profiles, which exhibit fewer localized features compared to other quantities. As Compton-y represents the overall thermal pressure integrated along the line of sight, it is a cumulative measurement. This integration leads to a smoothing effect, diminishing the sharp gradients or features observed in other profile types. As a result, the Compton-y profiles lack the detailed localized features that could be exploited for more precise parameter inference. Consequently, reducing the profile to a single integrated value leads to a smaller difference in the network's ability to infer the model parameters.

Previous studies have shown that when using \(Y_{200c}\), ASN2 was the best-inferred parameter among AAGN1, AAGN2, ASN1, and ASN2 \citep{2024ApJ...968...11L}. This result is reaffirmed here for \(Y_{200c}\) and extends to the results for \(T_{200c}\) and $\rho_g$, further supporting the robustness of these parameters.

Overall, these findings, while consistent with prior research, demonstrate that using the full profile generally offers a clear advantage in achieving higher accuracy in parameter inference across all physical quantities and parameters. This underscores the value of incorporating full profiles in astrophysical analyses, particularly for parameters where spatial variations within the cluster are crucial for accurate modeling and inference.


\section{Discussion and conclusions} 
\label{sec:conclusions}

In this study, we investigated the inference of cosmological and astrophysical parameters using averaged galaxy cluster profiles generated from the IllustrisTNG simulations. We performed parameter inference across the 28-dimensional IllustrisTNG model parameter space. The results show that stacked galaxy cluster profiles possess strong constraining power over this multidimensional parameter space, containing enough information to accurately infer all cosmological and astrophysical parameters of the model.

\begin{enumerate}

\item Stacked galaxy cluster profiles allow us to infer all cosmological and astrophysical parameters of the IllustrisTNG model with high accuracy in the noiseless case with 29 to 30 bins per profile. Indeed, including gas density, temperature, X-ray surface brightness, metallicity, and Compton-y profiles, we obtain coefficients approaching 0.97 for all cosmological parameters. For the remaining quantities, correlation coefficients remain above 0.90, reflecting very high accuracy across all parameters.

\item Our analysis reveals that different profile types exhibit varying sensitivities to specific parameters. For instance, temperature profiles generally provide the highest correlation coefficients for wind and feedback parameters, likely due to their sensitivity to energy injection processes within clusters. Gas density profiles also perform well, especially for cosmological parameters, suggesting that these profiles are highly responsive to the overall matter distribution within clusters. Conversely, the Compton-y profiles tend to show lower correlation coefficients, particularly for parameters related to localized processes, though their smoother nature may make them less susceptible to the specifics of subgrid models.

\item We observed that the inference accuracy for certain parameters diminishes as the cluster mass increases, which may be attributed to the self-similar nature of more massive clusters. These systems are less influenced by localized processes, such as star formation and feedback, leading to a reduced sensitivity to the parameters governing these processes. 

\item Radial cuts within clusters also reveal that the majority of the information necessary for accurate parameter inference is concentrated within the innermost regions, particularly within 0.1 \(R_{200,c}\). This finding underscores the importance of focusing on the central regions of clusters in future observational studies.

\item The addition of Gaussian noise to the profiles resulted in a general decline in correlation coefficients across all parameters, highlighting the sensitivity of the inference process to data quality. However, certain cosmological parameters, such as \(\Omega_{\rm m}\), \(\Omega_b\), and \(H_0\), along with specific astrophysical parameters like ASN2 and IMF, maintained relatively high correlation levels even under significant noise. These parameters also generate greater spread 
in the profiles, suggesting that both the non-degeneracy of the parameters and the strength of their influence on the profiles are key factors when dealing with noise. It is important to note that with a typical lower S/N value of 10, the inference remains highly accurate, with correlation coefficients exceeding 0.7 for these influential parameters. Even at an extremely low S/N ratio of 2.5, the impact on these parameters is minimal, indicating that their strong effects on the profiles make them more resistant to noise.

\item When comparing the inference capabilities of integrated quantities to full cluster profiles, our results show that full profiles generally provide a more accurate inference, particularly for parameters with significant spatial variations within the cluster. The integrated quantities, while consistent across different parameters, tend to underperform compared to the full profiles. This highlights the importance of considering the full spatial information contained within cluster profiles for precise parameter inference.

\end{enumerate}

Our findings indicate that stacked galaxy cluster profiles contain important information that allows us to effectively disentangle the effects of cosmology and astrophysical processes, providing accurate inference across a wide range of parameters. The robustness of our results across different mass bins, noise levels, and radial cuts suggests the potential of using stacked galaxy cluster profiles to maximize information extraction and enhance their utility for cosmological parameter inference. 

These results underscore the value of detailed profile analysis in astrophysical and cosmological research, particularly in the context of upcoming large-scale surveys. Future work could expand this approach to include more diverse simulation sets and even include observational data, further refining the accuracy and applicability of parameter inference in cosmology for the next-generation surveys.

\section*{Acknowledgments}

This work was supported by the grant agreements ANR-21-CE31- 0019/490702358 from the French Agence Nationale de la Recherche/DFG for the LOCALIZATION project. The CAMELS project is supported by NSF grants AST-2108944, AST-2108678, and AST-21080784.  The authors also thank the comments and discussions with all the members of the CAMELS project. The training of the MNNs has been carried out using graphics processing units (GPUs) from Simons Foundation, Flatiron Institute, Center of Computational Astrophysics.

\begin{appendix}
\section{Inference accuracy and dataset size}
\label{sec:A}

In this study, CARPoolGP's primary function is to generate a comprehensive dataset of mean galaxy cluster profiles for training, validating, and testing our neural network architecture. The optimal number of profiles needed to achieve accurate model performance while maintaining computational efficiency is not well-defined. To address this, we evaluated our neural network's performance using all five profile types together using different parameter space points:  300 (210 for training), 3,000 (2100 for training), 30,000 (21000 for training), and 300,000 (210000 for training). 

\begin{figure}
 \centering
 \includegraphics[scale=0.30]{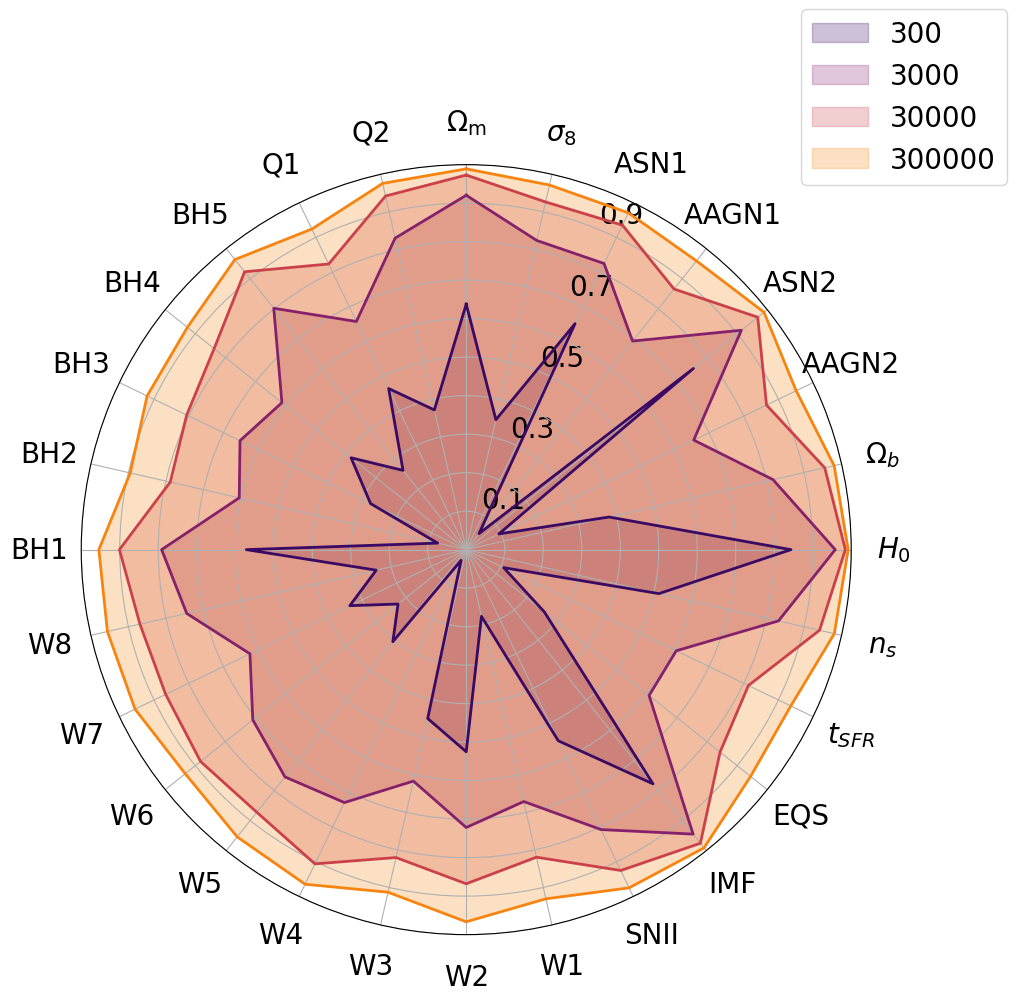}
     \caption{Radar chart displaying the correlation coefficients for various IllustrisTNG parameters as a function of input size. The radial axis represents the correlation coefficient values, with different symbols around the perimeter corresponding to specific IllustrisTNG parameters, including cosmological and astrophysical quantities. The shaded regions represent the results for different input sizes, ranging from 300 to 300,000 profiles. As the input size increases, the shaded regions expand outward, indicating improved correlation coefficients. Notably, the 30,000 profile input size approaches a saturation point, beyond which further increases in input size yield only marginal improvements in accuracy. This suggests that 30,000 profiles are sufficient to achieve robust parameter inference, with larger datasets providing only slight gains in accuracy across all parameters.}
\label{corr_coef_unlog_base_rad}
\end{figure}

Fig. \ref{corr_coef_unlog_base_rad} illustrates the correlation coefficients (\(r\)) for various IllustrisTNG parameters as a function of the input size used for model training, ranging from 300 to 300,000 parameter space locations. Notably, each input size involves the use of five distinct profiles per galaxy cluster—covering gas density, temperature, X-ray surface brightness, metallicity, and Compton-y parameters. The radar chart depicts how the accuracy of parameter inference improves with increasing input size, as indicated by the outward expansion of the shaded regions corresponding to each input size.

As the input size grows, there is a clear improvement in the correlation coefficients across all parameters. The analysis shows that with just 300 parameter space locations, the average correlation coefficient starts at a modest 0.5, but increases significantly to around 0.9 with 30,000 parameter space locations—reaching a point of apparent saturation. Beyond this threshold, moving to 300,000 parameter space locations yields only marginal gains, indicating diminishing returns with larger datasets. This suggests that using 30,000 parameter space locations with five profiles each is sufficient to capture most of the relevant information for robust parameter inference.

While the increased input size benefits most parameters, the extent of improvement varies. Cosmological parameters such as \(\Omega_{\rm m}\), \(\Omega_b\), and \(H_0\) maintain relatively stable and high correlation coefficients across all input sizes, highlighting their strong influence on cluster profiles. In contrast, some astrophysical parameters, especially those linked to complex feedback mechanisms (e.g., BH and AGN parameters), show more significant gains with larger input sizes, reflecting the greater complexity and variability in their effects on the profiles.

Certain parameters, like ASN2 and IMF, which are closely related to stellar content and feedback processes within clusters, exhibit high correlation coefficients even with smaller input sizes. This robustness suggests that these parameters exert a pronounced impact on the profiles, making them easier to infer accurately with fewer data points.

The findings from Fig. \ref{corr_coef_unlog_base_rad} have significant implications for our neural network training strategy. While using 300,000 parameter space locations may be beneficial for inference with a single profile type, the saturation point identified at around 30,000 parameter space locations suggests that training can be optimized by focusing on this input size when utilizing all five profile types together, achieving a balance between computational efficiency and inference accuracy. Consequently, we will use this number of profiles in the subsequent steps of our analysis.

\section{Profiles With One Parameter Variation}
\label{sec:B}

\begin{figure*}
 \centering
 \includegraphics[scale=0.198]{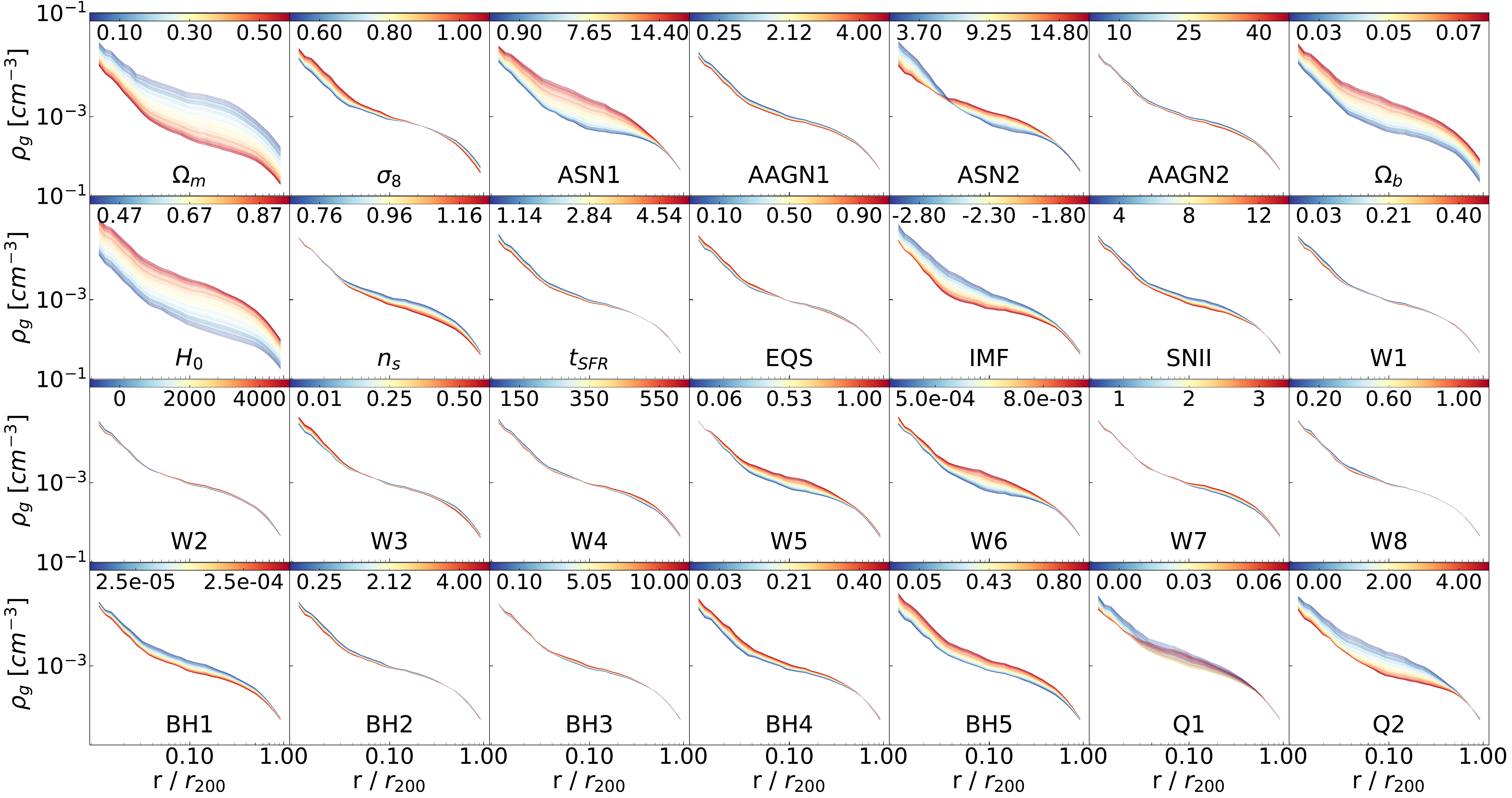}
     \caption{Gas density profiles showing the impact of varying individual parameters for all parameters in the IllustrisTNG model. Each line represents the profile obtained when a single parameter is varied while keeping the others fixed at their fiducial values. The color gradient within each panel highlights the spread introduced into the profiles by the variations in the corresponding cosmological parameter, with the color scale representing the range of parameter values. This visualization demonstrates how changes in cosmological parameters affect different physical quantities in galaxy clusters, providing insight into the sensitivity of cluster profiles to cosmological variations.}
\label{all_profiles_r200_ngas_test2.pdf}
\end{figure*}

\begin{figure*}
 \centering
 \includegraphics[scale=0.198]{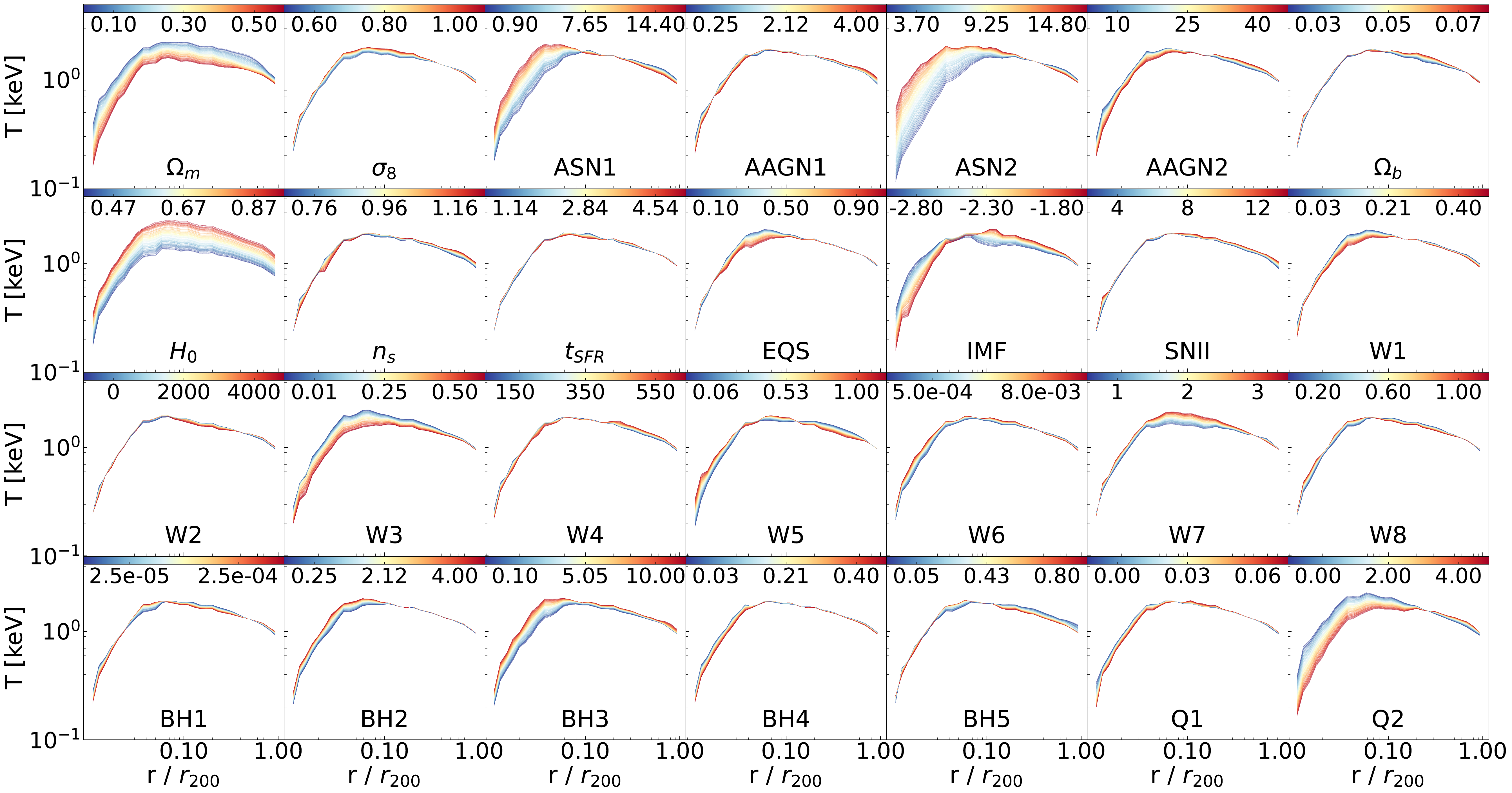}
     \caption{Similar to Figure \ref{all_profiles_r200_ngas_test2.pdf}, but for temperature profiles. Each line shows the effect of varying a single parameter while the others remain at their fiducial values, with the lines colored according to the value of the varying parameter.}
\label{all_profiles_r200_temp_test2.pdf}
\end{figure*}

\begin{figure*}
 \centering
 \includegraphics[scale=0.20]{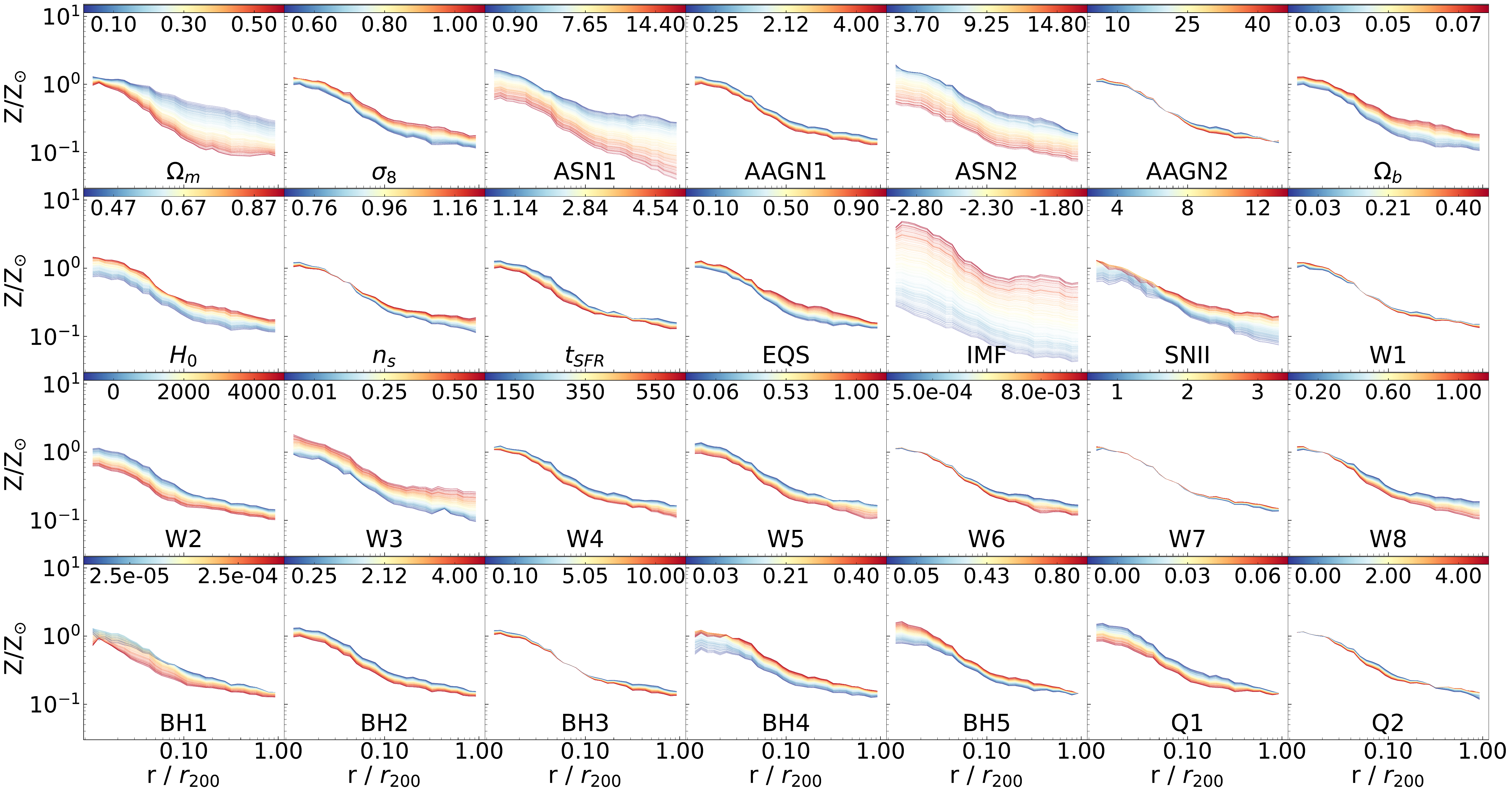}
     \caption{Similar to Figure \ref{all_profiles_r200_ngas_test2.pdf}, but for metallicity profiles. Each line shows the effect of varying a single parameter while the others remain at their fiducial values, with the lines colored according to the value of the varying parameter.}

\label{all_profiles_r200_zsun_test2.pdf}
\end{figure*}

\begin{figure*}
 \centering
 \includegraphics[scale=0.20]{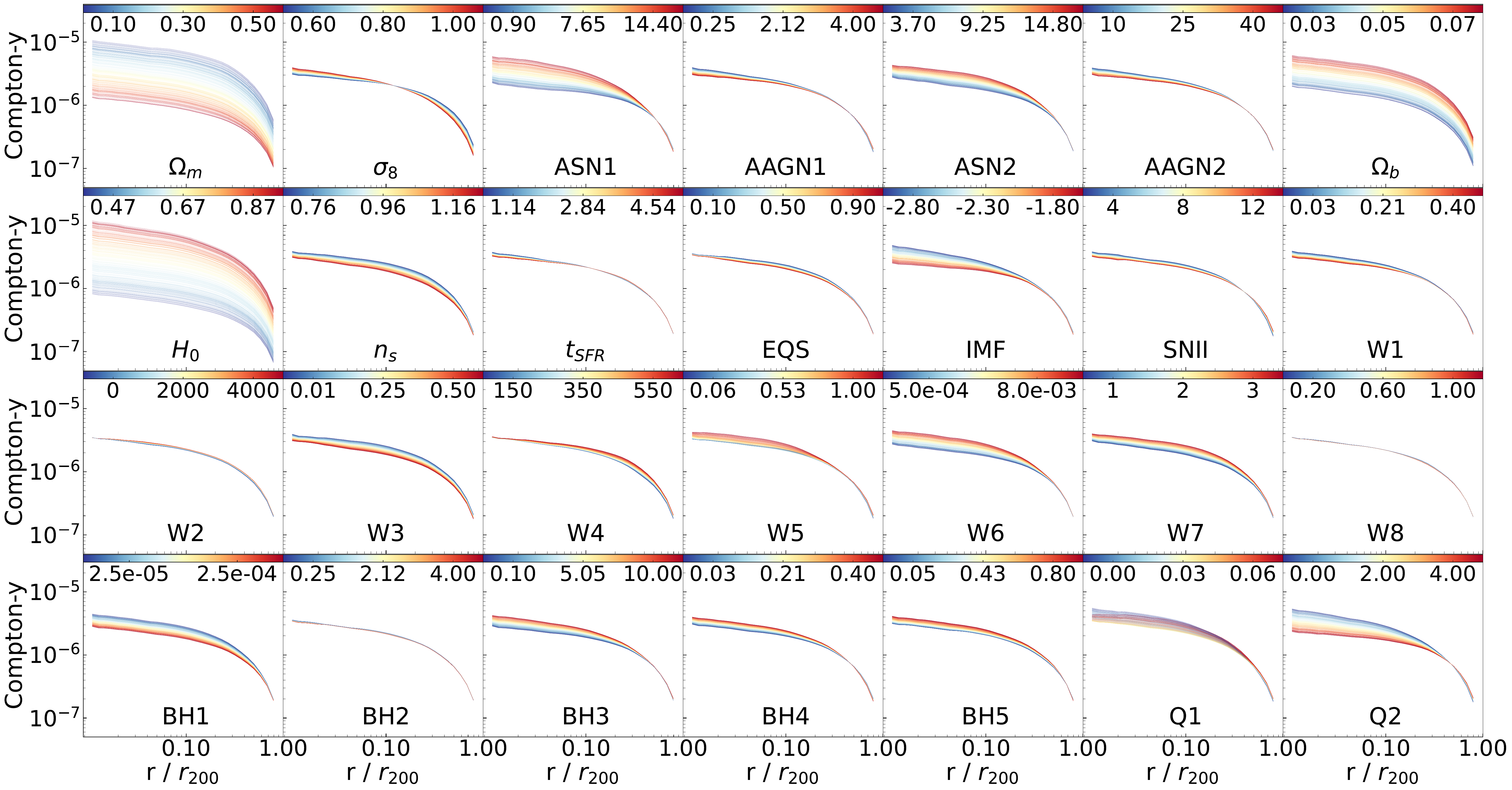}
     \caption{Similar to Figure \ref{all_profiles_r200_ngas_test2.pdf}, but for Compton-y profiles. Each line shows the effect of varying a single parameter while the others remain at their fiducial values, with the lines colored according to the value of the varying parameter.}
\label{all_profiles_r200_ycompt_test2.pdf}
\end{figure*}

\begin{figure*}
 \centering
 \includegraphics[scale=0.20]{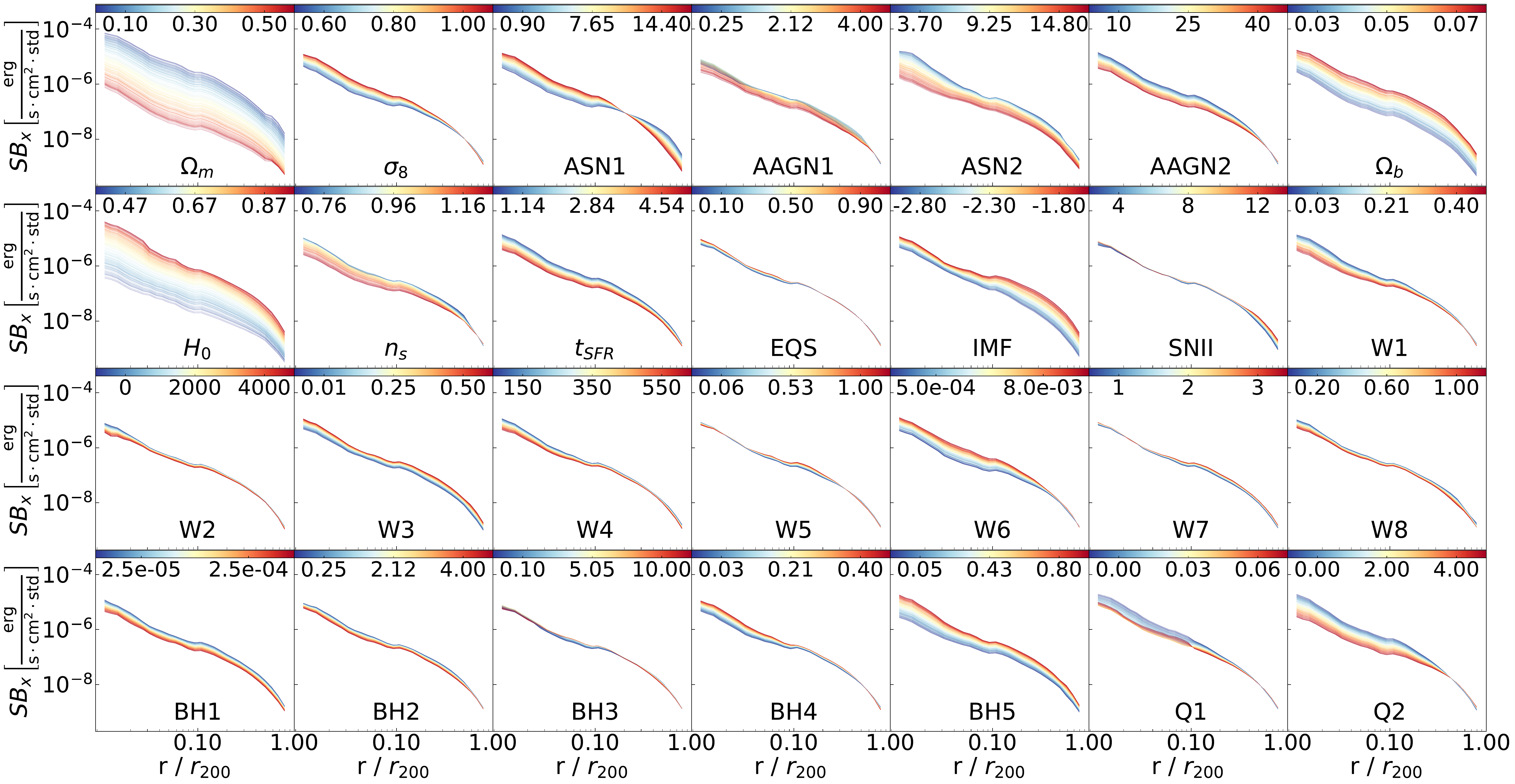}
     \caption{Similar to Figure \ref{all_profiles_r200_ngas_test2.pdf}, but for X-ray surface brightness profiles. Each line shows the effect of varying a single parameter while the others remain at their fiducial values, with the lines colored according to the value of the varying parameter.}
\label{all_profiles_r200_xray_test2.pdf}
\end{figure*}

To better understand the impact of individual parameters on the IllustrisTNG model on the averaged galaxy cluster profiles, we performed a series of emulations where we varied one parameter at a time while keeping the remaining 27 parameters fixed at their fiducial values. This method enabled a thorough exploration of each parameter's influence on the gas density, temperature, metallicity, Compton-y, and X-ray surface brightness mean profiles, as shown in Figs. 11-15. By isolating specific parameters, we gained deeper insights into their roles and identified potential degeneracies across different profile types.

We observe that cosmological parameters such as \(\Omega_{\rm m}\) and \(H_0\) result in strong spreads for all quantities, followed closely by \(\Omega_{b}\). Parameters such as \(\sigma_8\) and \(n_s\) show a more subtle effect on the profile variation for all quantities. 

Astrophysical parameters, while having a milder influence on the profiles for most quantities, show a stronger effect on metallicity, which is particularly sensitive to stellar parameters like the IMF and feedback and wind parameters such as  ASN1,  ASN2, and  W3.

The distinct responses of these profiles to the underlying parameters suggest that their combined effects are key to optimal parameter inference, helping to explain the model's exceptional performance.

A closer examination of Figs. 11-15 reveals that each parameter introduces distinct, unique features in the profiles across all quantities. This observation is key to understanding the high accuracy of our inference. These non-degenerate features enable the neural network to disentangle the individual effects of each parameter, allowing for parameter inference with an accuracy of 0.97 or higher in the noiseless case. This suggests that stacked galaxy cluster profiles contain clear and unambiguous information about both the astrophysical processes occurring in groups and clusters and the cosmological details of the Universe in which these structures reside.

\section{Consistency Tests}
\label{sec:C}

To verify that our neural network (NN) is not affected by any information leaks, we perform a series of tests.

\begin{figure}[ht!]
 \centering
 \includegraphics[scale=0.35]{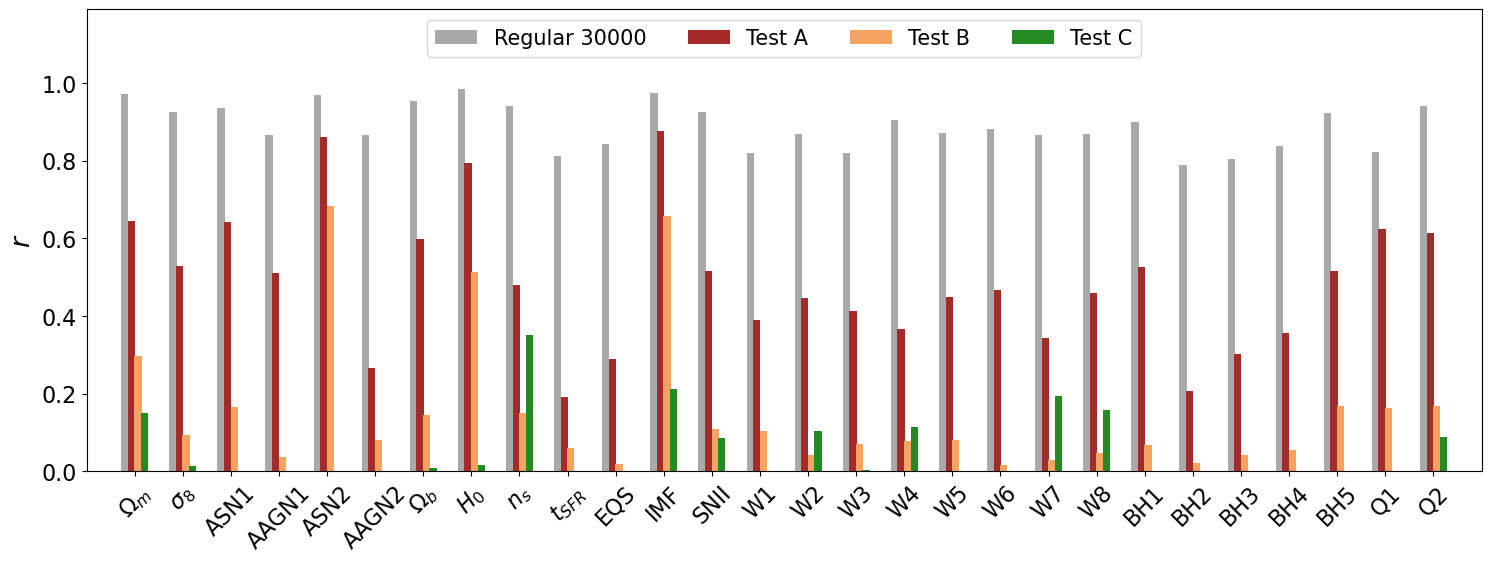}
     \caption{Correlation coefficients for the inference of 28 IllustrisTNG model parameters under two different scenarios. The grey bars represent the results using full galaxy cluster profiles with 30 bins per profile type and 30,000 parameter space locations, providing sufficient information to infer all parameters with high accuracy. The red bars show the results for Test A, where the number of input values is reduced to 25 (5 bins per profile type). As expected, the neural network fails to infer most parameters, accurately predicting only \(H_0\), the IMF slope, and ASN2. Test B is represented by the orange colored bars, where only one bin for each quantity (a total of 5 bins) was used, resulting in even lower correlation coefficients for all quantities. The green bars show the results of Test C, where the tetsing was performed on corrupted data, yielding correlation coefficients close to 0 for most parameters.}
\label{testA.png}
\end{figure}

\begin{figure*}
 \centering
 \includegraphics[scale=0.19]{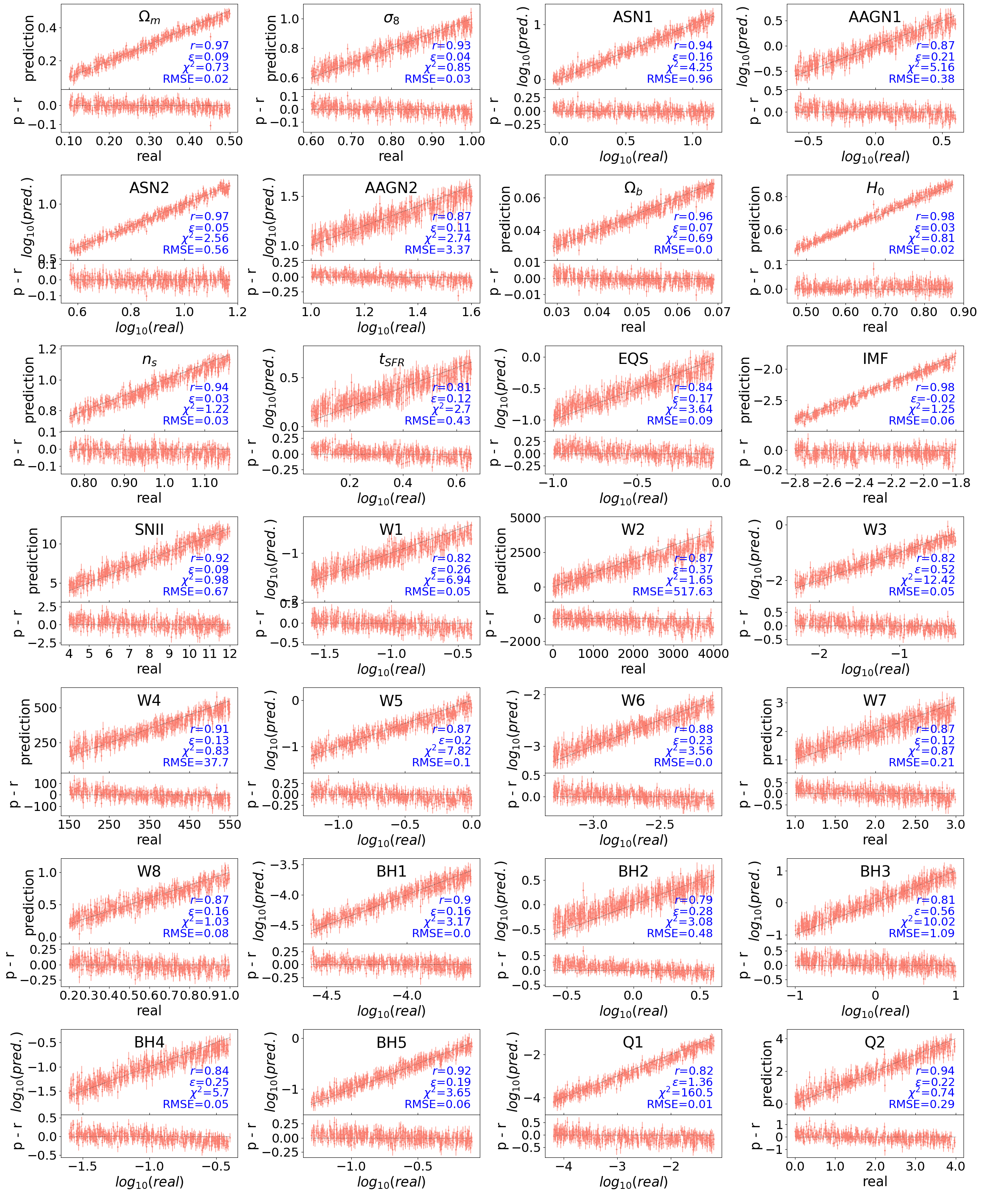}
     \caption{Inference results for all 28 IllustrisTNG model parameters using complete galaxy cluster profiles with full information across 30 bins. Each top panel shows the comparison between the true values (x-axis) and the predicted values (y-axis), with a black line indicating the one-to-one correspondence. The bottom of each panel displays the residuals to illustrate the spread. The correlation coefficient (\(r\)) and root-mean-square error (RMSE) mean relative error ($\epsilon$) and reduced chi-squared (\(\chi^2\)) are reported in blue for each parameter. These results demonstrate high accuracy across all parameters, with strong correlation coefficients and relatively low RMSE values. The accurate predictions underscore the ability of the model to infer both cosmological and astrophysical parameters when provided with full, detailed profiles, as reflected by the close alignment between predicted and true values and the small residuals across the parameter space.}
\label{all_30000.png}
\end{figure*}

\begin{figure*}
 \centering
 \includegraphics[scale=0.19]{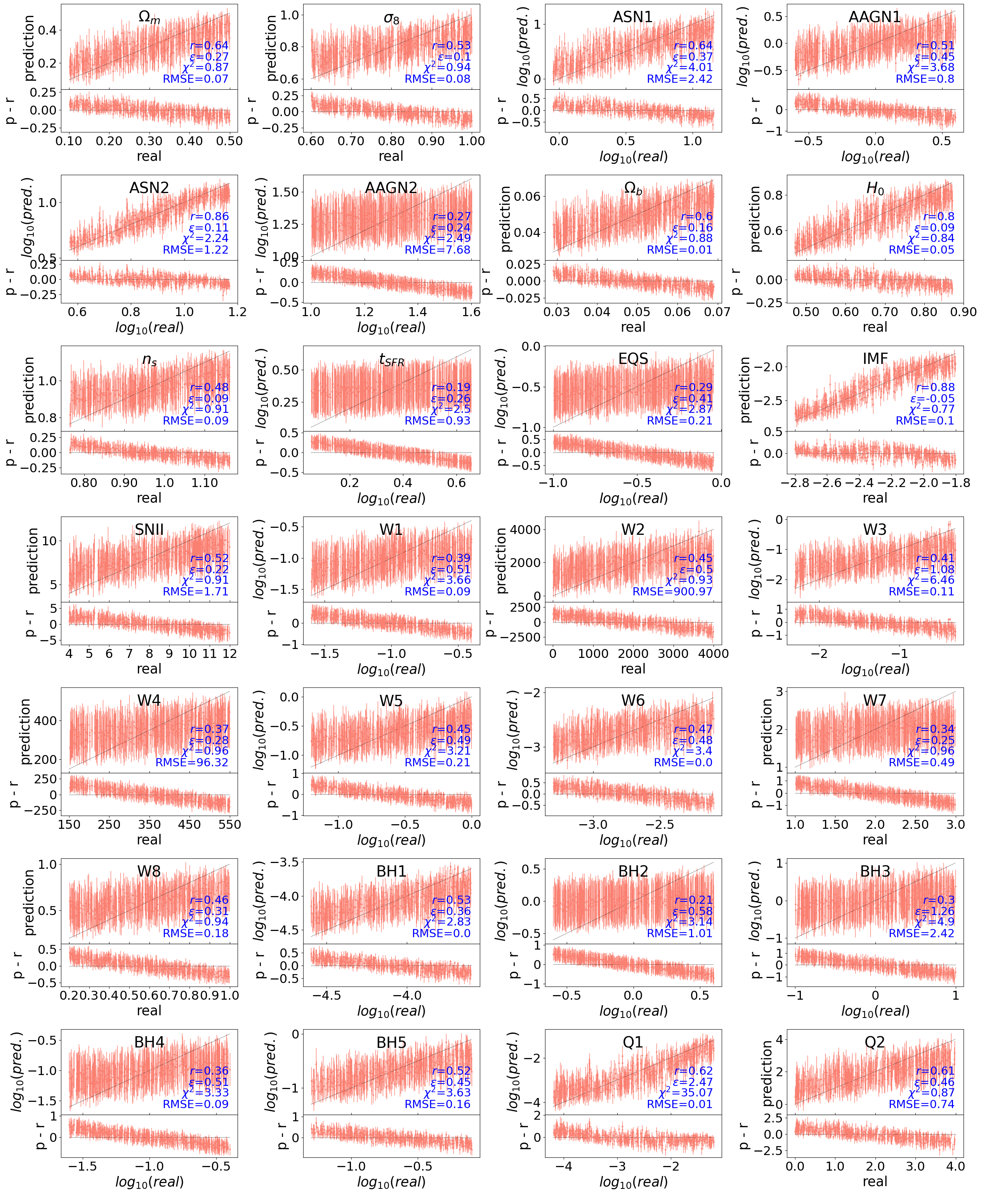}
     \caption{Inference results for test A. Similar to Fig. \ref{all_30000.png}, but here the results show the decrease in inference capabilities of the network when using an input size N=25. Only ASN2 and $H_0$ are infered with some moderate accuracy.}
\label{all_testA.png}
\end{figure*}

Test A uses 5 bins per quantity, resulting in 25 input values for each parameter space. Since it is not feasible to infer \(n\) parameters with fewer input values (\(N < n\)), we expect the NN to struggle when trying to infer 28 parameters from only 25 input values. We assess the NN's performance across 30,000 parameter space locations. The results, shown in Fig. \ref{testA.png}, compare the NN's performance with full galaxy cluster profiles (grey bars), which provide enough bins to infer all parameters, versus using only 25 input values (red bars). As expected, the NN fails to infer most parameters, successfully predicting only three—\(H_0\), the IMF slope, and ASN2—with moderate accuracy. Fig. \ref{all_testA.png} compares predicted and actual values, showing that while the trend is correct for these three parameters, there is more scatter and higher standard deviation compared to the results using full profiles (Fig. \ref{all_30000.png}). With insufficient input data, the network is unable to infer the remaining parameters.

\begin{figure*}
 \centering
 \includegraphics[scale=0.19]{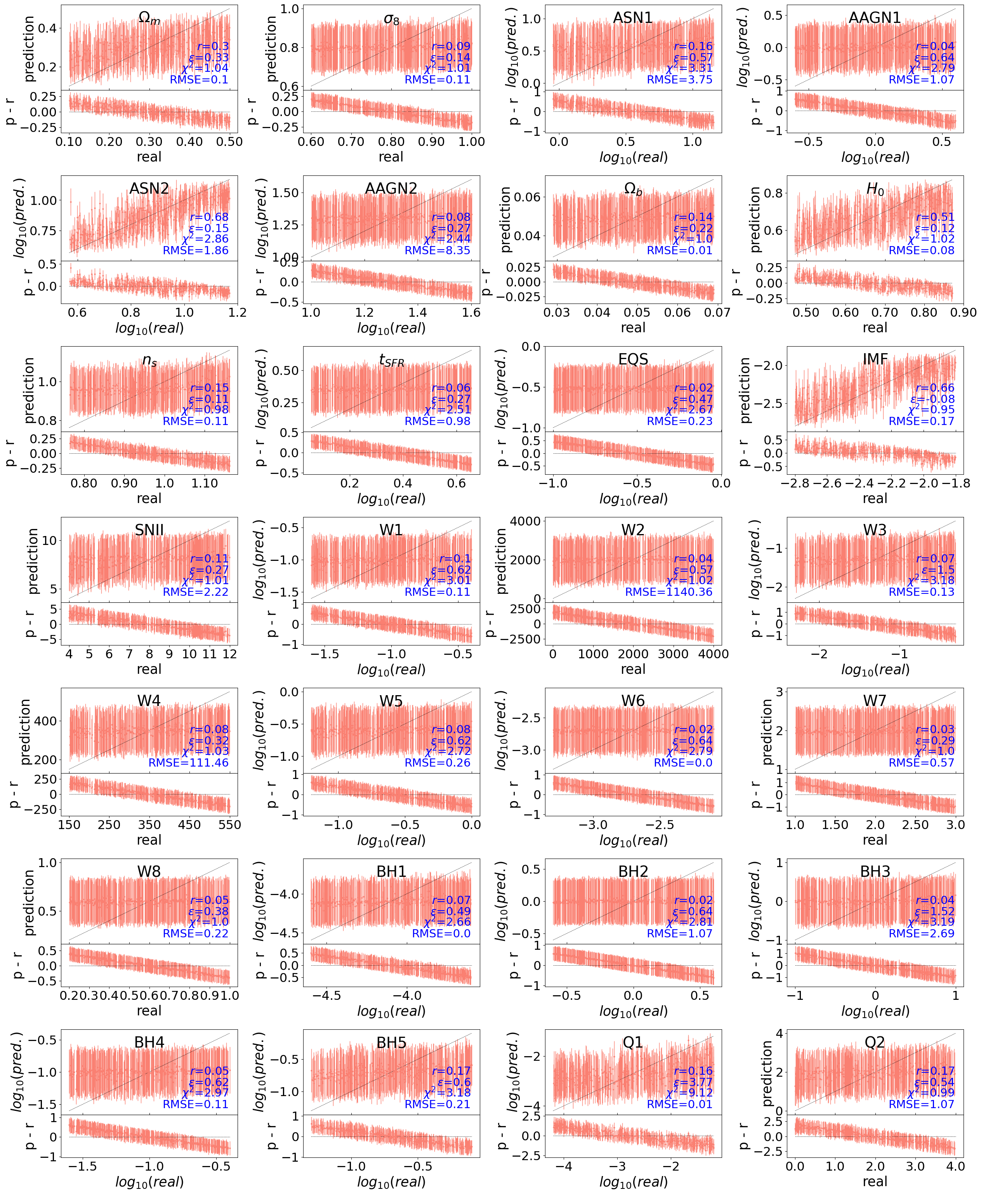}
     \caption{Inference results for test B. Similar to Fig. \ref{all_30000.png}, but here the results show how the NN is not capable of inferring any parameter when using an input size of N=5.}
\label{all_30000_testC.png}
\end{figure*}

Test B reduces the input vector size further by selecting only the first bin for each quantity, resulting in just 5 input values. The brown bars in Fig. \ref{testA.png} display the correlation coefficients, showing even further degradation in the NN's performance. Fig. \ref{all_30000_testC.png} compares predicted and actual values, highlighting that even for the best-inferred parameters such as \(H_0\), the IMF slope, and ASN2, scatter and uncertainty increase significantly, while the rest of the parameters are not inferred at all. 

\begin{figure*}
 \centering
 \includegraphics[scale=0.19]{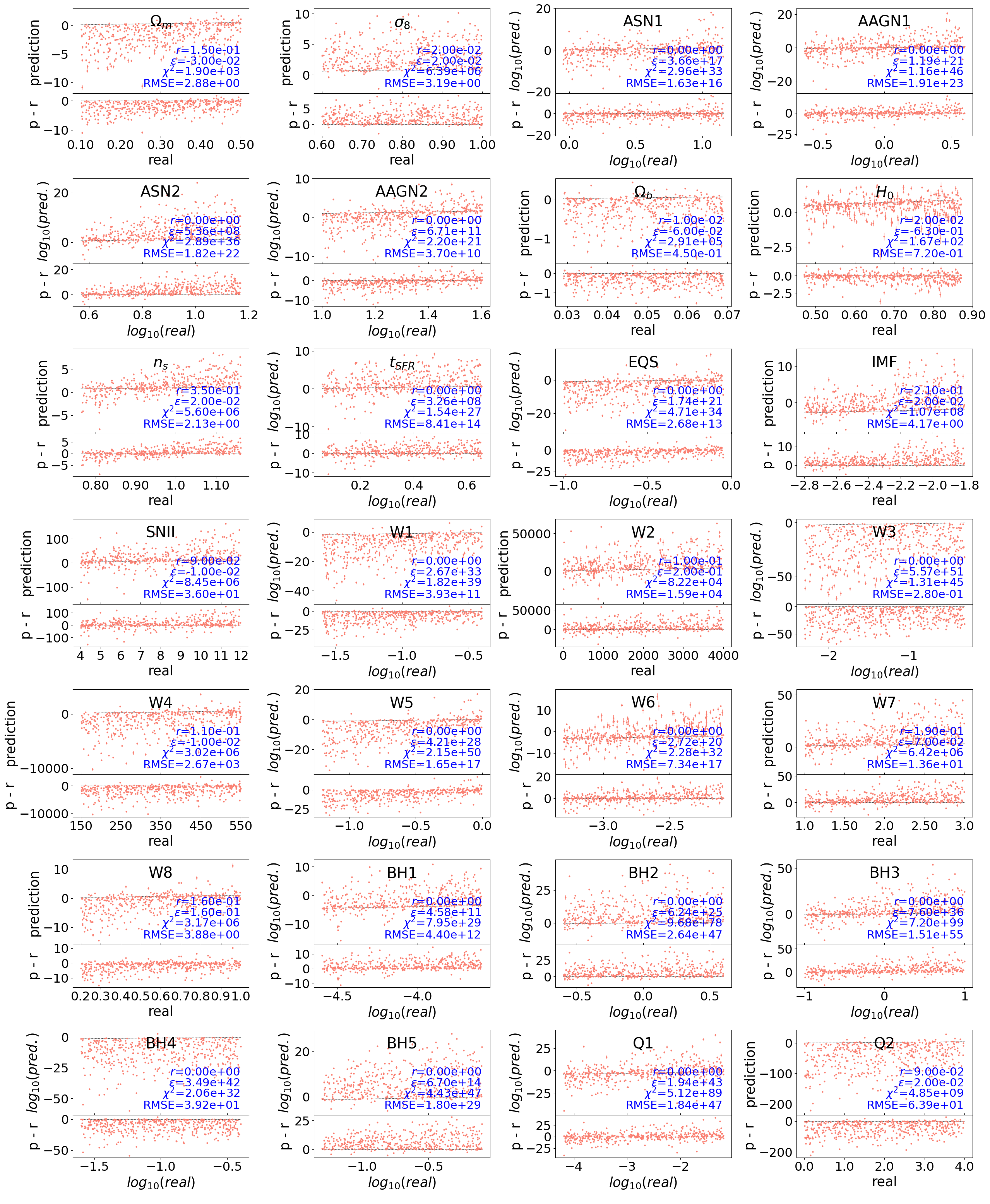}
     \caption{Inference results for test C. Similar to Fig. \ref{all_30000.png}. The results show that when using corrupted data for testing, the NN is not capable of inferring any of the parameters.}
\label{all_30000_testD.png} 
\end{figure*}

Test C examines the impact of corrupted data. Here, the network is trained using full profiles, but during testing, one profile is corrupted by adding random noise, effectively erasing its original values. The green bars in Fig. \ref{testA.png} display the correlation coefficients for this test, where the NN clearly fails to infer any parameters from the corrupted data. Fig. \ref{all_30000_testD.png} shows predicted values versus the mean for each quantity, demonstrating that the network completely misses its predictions when faced with corrupted data.

In conclusion, these tests confirm that our NN architecture does not suffer from information leaks, and the results in our study reflect physical phenomena rather than model artifacts.

\end{appendix}

\clearpage

\bibliography{sample631}{}
\bibliographystyle{aasjournal}

\end{document}